\documentclass[11pt,a4paper]{article}
\pdfoutput=1
\usepackage{amssymb,amsmath,amsfonts, mathtools}
\usepackage{graphicx}
\usepackage{caption}
\usepackage{subcaption}
\usepackage{enumerate}
\usepackage{dsfont}
\usepackage[sort,compress]{cite}
\usepackage{verbatim}
\usepackage{amsfonts,euscript,amssymb,stmaryrd,braket}
\usepackage{tikz}
\usetikzlibrary{arrows,decorations.markings,patterns}
\usepackage{slashed}
\definecolor{hyperref}{RGB}{026,028,185}
\usepackage[bookmarks=true,colorlinks=true,linkcolor=hyperref,citecolor=hyperref,urlcolor=hyperref,bookmarksnumbered]{hyperref}
\usepackage{cite}
\usepackage{gensymb}
 
\def\clock{{\count0=\time
           \divide\count0 60
           \ifnum\count0<10 0\fi\the\count0
           \multiply\count0 -60 \advance\count0 \time
           :\ifnum\count0<10 0\fi \the\count0
         }}
\newcommand{\timestamp}{{\small\vbox{\hbox{\tt\jobname.tex}
\hbox{\the\day/\the\month/\the\year, \clock}}}}

\newcommand{\nn}{\nonumber}
\newcommand{\ba}{\begin{eqnarray}}
\newcommand{\ea}{\end{eqnarray}}

\newcommand{\ads}{\textup{\textrm{AdS}}}

\newcommand{\sphere}{\textup{\textrm{S}}}

\newcommand{\be}{\begin{equation}}
\newcommand{\ee}{\end{equation}}

\makeatletter
\let\old@startsection=\@startsection
\let\oldl@section=\l@section
\renewcommand{\@startsection}[6]{\old@startsection{#1}{#2}{#3}{#4}{#5}{#6\mathversion{bold}}}
\renewcommand{\l@section}[2]{\oldl@section{\mathversion{bold}#1}{#2}}
\makeatother

\numberwithin{equation}{section}

\newcommand{\grp}[1]{\mathrm{#1}}

\newcommand{\grSO}{\grp{SO}}
\newcommand{\grSL}{\grp{SL}}

\newcommand{\p}{\partial}

\usepackage{color}

\def\s{\sigma}

\begin{document}
\renewcommand{\thefootnote}{\arabic{footnote}}

\overfullrule=0pt
\parskip=2pt
\parindent=12pt
\headheight=0in \headsep=0in \topmargin=0in \oddsidemargin=0in

\vspace{ -3cm} \thispagestyle{empty} \vspace{-1cm}
\begin{flushright} 
\footnotesize
HU-EP-15/58\\
\end{flushright}%

\begin{center}
\vspace{1.2cm}
{\Large\bf \mathversion{bold}
Precision calculation of 1/4-BPS Wilson loops in AdS$_5\times \sphere^5$

}

 \vspace{0.8cm} {
  V.~Forini$^{a,}$\footnote{ {\tt $\{$valentina.forini,edoardo.vescovi$\}$@\,physik.hu-berlin.de}},
 V.~Giangreco M. Puletti$^{b,}$\footnote{ {\tt vgmp@hi.is}},
  L.~Griguolo$^{c,}$\footnote{ {\tt luca.griguolo@fis.unipr.it}}, 
D.~Seminara$^{d,}$\footnote{ {\tt seminara@fi.infn.it}},
E.~Vescovi$^{a,1}$}
 \vskip  0.5cm

\small
{\em
$^{a}$Institut f\"ur Physik, Humboldt-Universit\"at zu Berlin, IRIS Adlershof, \\Zum Gro\ss en Windkanal 6, 12489 Berlin, Germany  
\vskip 0.05cm
$^{b}$ University of Iceland,
Science Institute,
Dunhaga 3,  107 Reykjavik, Iceland
  \vskip 0.05cm
$^{c}$ Dipartimento di Fisica e Scienze della Terra, Universit\'a di Parma and INFN
Gruppo Collegato di Parma, Viale G.P. Usberti 7/A, 43100 Parma, Italy
  \vskip 0.05cm
$^{d}$ Dipartimento di Fisica, Universit\'a di Firenze and INFN Sezione di Firenze, Via G. Sansone 1, 50019 Sesto Fiorentino,
Italy
}
\normalsize

\end{center}

\vspace{0.3cm}
\begin{abstract} 
\noindent We study the strong coupling behaviour of $1/4$-BPS circular Wilson loops (a family of ``latitudes")
in ${\cal N}=4$ Super Yang-Mills theory, computing the one-loop corrections
to the relevant classical string solutions in AdS$_5\times$S$^5$. Supersymmetric localization provides an exact result that, in the large 't Hooft coupling limit, should be reproduced by the sigma-model approach. To avoid ambiguities due to the absolute normalization of the string partition function, we compare the $ratio$ between the generic latitude and the maximal 1/2-BPS circle:  Any measure-related ambiguity should simply cancel in this way. We use the Gel'fand-Yaglom method with Dirichlet boundary conditions to calculate the relevant functional determinants, that present some complications with respect to the standard circular case. After a careful numerical evaluation of our final expression we still find disagreement with the localization answer: The difference is  encoded into a precise ``remainder function". We comment on the possible
origin and resolution of this discordance.

\end{abstract}

\newpage

\tableofcontents

\newpage

\section{Introduction and main result}
\label{sec:intro}

The harmony between exact QFT results obtained through localization procedure for BPS-protected Wilson loops in $\mathcal{N}=4$ SYM and their stringy counterpart is a thorny issue beyond the supergravity approximation. 
For the $1/2$-BPS circular Wilson loop~\cite{Berenstein:1998ij,Drukker:1999zq}, in the fundamental representation,  supersymmetric localization~\cite{Pestun:2007rz} in the gauge theory  confirms the all-loop  prediction based on a large $N$ resummation of ladder Feynman diagrams~\cite{Erickson:2000af}~and generalized to finite $N$ in~\cite{Drukker:2000rr}. 
On the string theory side, this should equate 
the disc partition function for the $\ads_5\times \sphere^5$ superstring. 
Its one-loop contribution, encoding fluctuations above the classical solution,  has been formally written down in~\cite{Drukker:2000ep},  explicitly evaluated in~\cite{Kruczenski:2008zk}~\footnote{See also~\cite{Sakaguchi:2007ea}.} using the Gel'fand-Yaglom method, reconsidered in~\cite{Kristjansen:2012nz} with a different choice of boundary conditions and reproduced in~\cite{Buchbinder:2014nia}~\footnote{See Appendix B in~\cite{Buchbinder:2014nia}.} with the heat-kernel technique.  
No agreement was found with the subleading correction in the strong coupling ($\lambda\gg1$) expansion of the gauge theory result in the planar limit
\be\label{circlevev}
\log\langle\mathcal{W}\left(\lambda,\theta_0=0\right)\rangle=\log{\textstyle\frac{2}{\sqrt{\lambda}}}I_1(\sqrt{\lambda}) 
=\sqrt{\lambda}-\frac{3}{4}\,\log\lambda+\frac{1}{2}\,\log\frac{2}{\pi}+\mathcal{O}(\lambda^{-\frac{1}{2}})~,
\ee
where $I_1$ is the modified Bessel function of the first kind, the meaning of the parameter $\theta_0$ is clarified below, and the term proportional to $\log\lambda$ in \eqref{circlevev}  is argued to originate from the $\grSL(2,\mathbb{R})$ ghost zero modes on the disc~\cite{Drukker:2000rr}. The discrepancy occurs  in the $\lambda$-independent  part above~\footnote{See formula \eqref{kructirziu} below.}, originating from  the one-loop effective action contribution \emph{and} an unknown, overall numerical factor in the measure of the partition function. 
 
The situation becomes even worse when considering a loop winding $n$-times around itself \cite{Kruczenski:2008zk,Bergamin:2015vxa}, where also the functional dependence on $n$ is failed by the one-loop string computation. The case of different group representations has also been considered: For the $k$-symmetric and $k$-antisymmetric representations, whose gravitational description is given in terms of D3- and D5-branes, respectively, the first stringy correction again does not match the localization result \cite{Faraggi:2014tna}. Interestingly, the Bremsstrahlung function of ${\cal N}=4$ SYM, derived in \cite{Correa:2012at} again using a localization procedure, is instead correctly reproduced \cite{Drukker:2011za} through a one-loop computation around the classical cusp solution \cite{Drukker:1999zq,Drukker:2007qr}.

Localization has been proven to be one of the most powerful tools in obtaining non perturbative results in quantum supersymmetric gauge theories~\cite{Pestun:2007rz}: An impressive number of new exact results have been derived in different dimensions, mainly when formulated on spheres or products thereof~\cite{Pestun:2007rz, Kapustin:2009kz}. In order to gain further intuition on the relation between localization and sigma-model perturbation theory in different and more general settings, we re-examine this issue addressing as follows the problem of how to possibly eliminate the ambiguity related to the partition function measure. We consider the string dual to a non-maximal circular Wilson loop - the family of 1/4-BPS operators with path corresponding to a latitude in $\sphere^2\in \sphere^5$ parameterized by an angle $\theta_0$ and studied at length in~\cite{Drukker:2005cu,Drukker:2006ga,Drukker:2007qr} - and evaluate the corresponding string one-loop path integral. We then calculate the \emph{ratio} between the latter and the corresponding one representing the maximal circle - the case $\theta_0=0$ in \eqref{circlevev}. 
 Our underlying  assumption is that the measure is actually independent on the geometry of the worldsheet associated to the Wilson loop~\footnote{
About the \emph{topological} contribution of the measure, its relevance in canceling the divergences occurring in evaluating quantum corrections to the string partition function has been first discussed in~\cite{Drukker:2000ep} after the observations of~\cite{Forste:1999qn,Forste:1999cp}.
We use this general argument  below, see discussion around \eqref{Eulervolume}. }, and therefore in such ratio measure-related ambiguities should simply cancel. It appears non-trivial to actually prove a background independence of the measure, whose diffeo-invariant  definition includes in fact explicitly the worldsheet fields~\footnote{See for example the discussion in~\cite{Roiban:2007jf}.}. 
Our assumption -- also suggested in~\cite{Kruczenski:2008zk} -- seems however a reasonable one, especially in light of  the absence of zero mode in the classical solutions here considered~\footnote{In presence of zero mode, a possible dependence of the path integral measure on the classical solution comes from the integration over collective coordinates associated to them. In this framework, see discussion in \cite{Zarembo:2002an}.} and of the explicit example of (string dual to)  the ratio of 
a cusped Wilson loop with a straight line~\cite{Drukker:2011za}, where a
perfect agreement exists between sigma model perturbation theory and localization/integrability results~\cite{Correa:2012at}~\footnote{See also~\cite{Forini:2010ek}, which analyzes the (string dual to the) ratio between the Wilson loop of  ``antiparallel lines'' and straight line.}.
  
The family of 1/4-BPS latitude Wilson loops falls under the more general class of 1/8-BPS Wilson loops with arbitrary shape on a two-sphere introduced in~\cite{Drukker:2007dw,Drukker:2007yx,Drukker:2007qr} and studied in ~\cite{Pestun:2009nn}. There are strong evidences that they  localize into Yang-Mills theory on $\sphere^2$ in the zero-instanton sector \cite{Pestun:2009nn,Young:2008ed,Bassetto:2008yf,Drukker:2007qr} and their vacuum expectation values are therefore related to the 1/2-BPS one by a simple rescaling. As originally argued in~\cite{Drukker:2006ga} the expectation value of such latitude Wilson loops is obtained from the one of the maximal circle provided one replaces $\lambda$ with an effective 't Hooft coupling $\lambda'=\lambda \cos^2\theta_0$. The ratio of interest follows very easily
\begin{gather}\label{mainratio}
\frac{\langle\mathcal{W}\left(\lambda,\theta_0\right)\rangle}{\langle\mathcal{W}\left(\lambda,0\right)\rangle}\biggr\rvert_{\rm loc}=e^{\sqrt{\lambda}\left(\cos\theta_0-1\right)}\left[(\cos\theta_0)^{-\frac{3}{2}}+\mathcal{O} (\lambda^{-\frac{1}{2}} )\right]
+\mathcal{O}\left(e^{-\sqrt{\lambda}}\right)\,,
\end{gather} 
where in the large $\lambda$ expansion only the dominant exponential contribution is kept (and $\rm loc$ stands for ``localization''). In terms of string one-loop effective actions $\Gamma=-\log Z\equiv -\log\langle W\rangle$,  this leads to the prediction
\begin{gather}\label{mainratiolog}
 \log\frac{\langle\mathcal{W}\left(\lambda,\theta_0\right)\rangle}{\langle\mathcal{W}\left(\lambda,0\right)\rangle}\biggr\rvert_{\rm loc}=\left[ \Gamma(\theta_0=0)-\Gamma(\theta_0)\right]_{\textrm{loc}}=\sqrt{\lambda}\,(\cos\theta_0-1 )-\frac{3}{2}\log\cos\theta_0+\mathcal{O} (\lambda^{-\frac{1}{2}} ) ~,
\end{gather} 
where the leading term comes from the regularized minimal-area surface of the strings dual to these Wilson loops, while the semiclassical string fluctuations in the  string sigma-model account for the subleading correction.

As usual, the one-loop contribution derives from the evaluation of ratios of functional determinants in the quadratic expansion of the type IIB Green-Schwarz action about the string classical background. The axial symmetry of the worldsheet surface simplifies these two-dimensional spectral problem to infinitely-many one-dimensional spectral problems. To solve them, we use the Gel'fand-Yaglom method originally developed in \cite{Gelfand:1959nq} and later improved in a series of papers \cite{Forman1987, Forman1992, McKane:1995vp, Kirsten:2003py, Kirsten:2004qv, Kirsten:2007ev}. A concise review of this technique is presented in Appendix \ref{app:gelfand_yaglom}. Unlike other procedures ({\it e.g.} heat kernel \cite{Bergamin:2015vxa}), this method of regularizing determinants effectively  introduces a fictitious boundary for the worldsheet surface, besides the expected conformal one.
We then proceed with the analytical computation of the functional determinants by imposing Dirichlet boundary conditions on the bosonic and fermionic fluctuation fields at the conformal (AdS boundary) and fictitious boundaries, whose contribution effectively vanish in the chosen regularization scheme \cite{Frolov:2004bh, Dekel:2013kwa}. We emphasise that this procedure differs from the one employed in \cite{Kruczenski:2008zk}, since the non-diagonal matrix structure of the fermionic-fluctuation operator for arbitrary $\theta_0$ prevents us from factorizing the value of the fermionic determinants into a product of two contributions.

In the $\theta_0\to 0 $ limit, we analytically recover the constant one-loop coefficient in the expansion of the 1/2-BPS circular Wilson loop as found in \cite{Kruczenski:2008zk,Buchbinder:2014nia}
\be\label{kructirziu}
\log\langle\mathcal{W}\left(\lambda,\theta_0=0\right)\rangle
=\sqrt{\lambda}-\frac{3}{4}\,\log (\lambda)+\log c+\frac{1}{2}\,\log\frac{1}{2\pi}+\mathcal{O}(\lambda^{-\frac{1}{2}})~,
\ee
up to an unknown contribution of ghost zero-modes (the constant $c$). The expression above is in disagreement with the gauge theory prediction \eqref{circlevev}.

We regularize and normalize the latitude Wilson loop with respect to the circular case. 
The summation of the one-dimensional Gel'fand-Yaglom determinants is quite difficult, due to the appearance of some Lerch-type special functions, and we were not able to obtain a direct analytic result. We resort therefore to a numerical approach. Our analysis shows that the disagreement between sigma-model and localization results \eqref{mainratiolog} is not washed out yet. Within a certain numerically accuracy, we claim that the discovered $\theta_0$-dependent discrepancy is very well quantified as 
\begin{eqnarray}
 \log\frac{\langle\mathcal{W}\left(\lambda,\theta_0\right)\rangle}{\langle\mathcal{W}\left(\lambda,0\right)\rangle}\biggr\rvert_{\rm sm}
=
\sqrt{\lambda}\,(\cos\theta_0-1 )-\frac{3}{2}\log\cos\theta_0+\log \cos\frac{\theta_0}{2}
+\mathcal{O} (\lambda^{-\frac{1}{2}} ) ~,
\end{eqnarray}
suggesting that the ``remainder function" should be 
\be
{\rm Rem }(\theta_0)=\log \cos\frac{\theta_0}{2}.
\ee

\bigskip


Before proceeding with the numerical analysis a series of non-trivial steps have been performed and the final expression appears as the result of precise cancellations. As already remarked, the fermionic determinants do not trivially factorize, and consequently we have to solve a coupled Schr\"{o}dinger  system to deal with the Gel'fand-Yaglom method. 
It turns out that the decoupling of the fictitious boundary relies on  delicate compensations between bosonic and fermionic contributions, involving different terms of the sum. As a matter of fact, after removing the infrared regulator  we obtain the correct ultraviolet divergencies from the resulting effective actions. These are subtracted from the final ratio in order to obtain a well-behaved sum, amenable of a numerical treatment. Unfortunately the final result is inconsistent with the QFT analysis, opening the possibility that something subtle is missing in our procedure. On the other hand we think that our investigation elucidates several points at least in the standard setup to solve the spectral problem,  and thus should be helpful for further developments. We will comment on the possible origin  of the discrepancy at the end of the manuscript.  
 
\bigskip

The paper proceeds as follows. In Section \ref{sec:classical} we recall  the classical setting, in Section \ref{sec:determinants} we evaluate the relevant functional determinants which we collect in Section \ref{sec:partitionfunctions} to form the corresponding partition functions. Section \ref{sec:conclusions}  contains concluding remarks on the disagreement with the localization result and its desirable explanation. After a comment on notation in Appendix   \ref{app:notation}, we devote Appendix \ref{app:gelfand_yaglom} to a concise survey on the Gel'fand -Yaglom method. Appendix \ref{app:detailsferm} elucidates some properties which simplify the evaluation of the fermionic contribution to the partition function, while in Appendix \ref{bclower} we comment on a possible different choice of boundary condition for lower Fourier modes which that not affect our results.


\section{Classical string solutions dual to latitude Wilson loops}
\label{sec:classical}

The classical string surface describing  the strong coupling regime of the  $1/4$-BPS {\it latitude} was  first found  in~\cite{Drukker:2005cu} and discussed in details in \cite{Drukker:2006ga,Drukker:2007qr}. Endowing the $\ads_5\times \sphere^5$ space with a Lorentzian metric in global coordinates
\begin{eqnarray}
  ds^2_{\textrm{10D}}  & = & -\cosh^2\rho dt^2+d\rho^2+\sinh^2\rho
  \left( d\chi^2 +\cos^2\chi d\psi^2+\sin^2\chi d\varphi_1^2 \right)\nonumber\\
  && +d\theta^2+\sin^2\theta d\phi^2+\cos^2\theta
  \left(d\vartheta_1^2+\sin^2\vartheta_1\left(d\vartheta_2^2+\sin^2\vartheta_2 d\varphi_2^2\right) 
  \right)\,,
  \label{metric_old}
\end{eqnarray}
with the AdS radius set to 1, the corresponding  classical configuration  in $\ads_3\times \sphere^2$
\begin{equation}
  \label{background_old}
  \begin{split}
 t&=0, \qquad \rho=\rho(\sigma), \qquad \chi=0, \qquad \psi=\tau, \qquad \qquad \varphi_1=\text{const}, \\
\theta&=\theta(\sigma), \qquad \phi=\tau, \qquad \vartheta_1=0, \qquad \vartheta_2=\text{const}, \qquad \varphi_2=\text{const},
 \end{split}
 \end{equation}
parametrizes a string worldsheet, ending on a unit circle at the boundary 
 of $\ads_5$ and on a latitude sitting at polar angle $\theta_0$ on a two-sphere inside the compact space 
 \footnote{There exist other solutions with more wrapping in $\sphere^5$, but they are not supersymmetric~\cite{Drukker:2006ga}.}. 
Here the polar angle $\theta$ spans  the interval $[-\frac{\pi}{2},\frac{\pi}{2}].$ 
 The worldsheet coordinates instead take values in the range $\tau\in [0,2\pi)$ and 
$\sigma\in [0,\infty)$. \\
The  ansatz \eqref{background_old}  does not propagate along the time direction and defines an Euclidean surface embedded in a Lorentzian target space. It
satisfies the equation of motions (supplemented by the Virasoro constraints in the Polyakov formulation) when we set
\begin{equation}   \label{rho_and_theta}
\begin{split}
 &  \sinh\rho(\sigma)=\frac{1}{\sinh\sigma},\qquad \qquad~
   \cosh\rho(\sigma)=\frac{1}{\tanh\sigma}, \\
&   \sin\theta(\sigma)=\frac{1}{\cosh\left(\sigma_0\pm\sigma\right)}, 
   \qquad
   \cos\theta(\sigma)=\tanh\left(\sigma_0\pm\sigma\right).
\end{split}
\end{equation}
An integration constant in \eqref{rho_and_theta} that shifts $\sigma$ was chosen to be zero so that the worldsheet boundary at 
$\sigma=0$ is located at the boundary of $\ads_5$.
The remaining one, $\sigma_0\in[0,\infty)$, spans the one-parameter family of latitudes on $\sphere^5$ at the boundary 
$\sigma=0$, whose angular position $\theta_0\in[0,\frac{\pi}{2}]$ relates to $\sigma_0$
through
\begin{gather}
  \label{theta_0_and_sigma_0}
  \cos\theta_0=\tanh\sigma_0.
\end{gather}
Here the dual gauge theory operator interpolates between two notable cases. The 1/2-BPS circular case falls under this class of Wilson loops when the latitude in $\sphere^2$ shrinks to a point for $\theta_0=0$, which implies $\theta(\sigma)=0$ and  $\sigma_0=+\infty$ from 
(\ref{rho_and_theta})-(\ref{theta_0_and_sigma_0}). In this case the string propagates only in $\ads_3$. The other case is the circular 1/4-BPS Zarembo Wilson loop when the worldsheet extends over a maximal circle of $\sphere^2$ for $\theta_0=\frac{\pi}{2}$ and $\sigma_0=0$ \cite{Zarembo:2002an}~\footnote{
See also~\cite{Miwa:2015bta} for an analysis of the contribution to the string partition function due to (broken) zero modes of the solution  in~\cite{Zarembo:2002an}.}. The double sign in \eqref{rho_and_theta} accounts for the existence of two solutions, effectively doubling the range of $\theta_0$:
The stable (unstable) configuration mimizes (maximizes) the action functional and wraps the north pole $\theta=0$ (south pole $\theta=\pi$) of $\sphere^5$. \\
The semiclassical analysis is more conveniently carried out in the stereographic coordinates
$\upsilon^m$ ($m=1,2,3$) of $\sphere^3\subset \ads_5$ and $w^n$ ($n=1,2,3,4,5$) of $\sphere^5$ 
\begin{gather} \label{metric_new}
  ds^2_{\textrm{10D}}  =  -\cosh^2\rho dt^2+d\rho^2+\sinh^2\rho\,
  \frac{d\upsilon_m d\upsilon_m}{ (1+\frac{\upsilon^2}{4} )^2} +\frac{d w_n d w_n}{ (1+\frac{w^2}{4} )^2}\,,\\
   \upsilon^2= \upsilon_m \upsilon_m \qquad w^2= w_n  w_n
\end{gather}
where the classical solution reads~\footnote{The background of $\varphi_1,\varphi_2,\vartheta_2$ was set to zero in \eqref{background_old}, but the bosonic quadratic Lagrangian does not have the standard form (kinetic and mass terms for the eight physical fields) in the initial angular coordinates.}
\begin{equation}
 \label{background_new}
\begin{split}
 t&=0, \qquad\quad~~~ \rho=\rho(\sigma), \qquad ~~~ \upsilon_1=2\sin\tau, \qquad~~~  \upsilon_2=2\cos\tau, 
 \qquad~~~  \upsilon_3=0\,,\\
  w_1&=w_2=0, \qquad w_3=2\cos\theta(\sigma), \qquad 
 w_4=2\sin\theta(\sigma)\sin\tau, \qquad 
 w_5=2\sin\theta(\sigma)\cos\tau\, .
 \end{split}
 \end{equation} 
The induced metric on the worldsheet depends on the latitude angle $\theta_0$ through the conformal factor ($\sigma^i=(\tau,\sigma)$)\begin{gather}\label{induced_metric}
  ds^2_\textrm{2D}
  =h_{ij}d\sigma^i d\sigma^j
  =\Omega^2(\sigma) \left(d\tau^2+d\sigma^2\right)\,,\qquad \Omega^2(\sigma)\equiv\sinh^2\rho(\sigma)+\sin^2\theta(\sigma)\, .
\end{gather}
The two-dimensional Ricci curvature is then
\begin{align}\label{R2}
^{(2)}\!\!\, R  =&-\frac{2\,\partial_\sigma^2\log\Omega(\s)}{\Omega^2(\s)}=\\
=&\mbox{\small $\displaystyle -\frac{ \left(2 \cosh 2 \sigma _0\pm 2 \sinh  \sigma _0 \sinh \left(6 \sigma\pm3 
   \sigma _0 \right)-3 \cosh
   \left(2 \left(\sigma\pm \sigma _0 \right)\right)+6 \cosh \left(4 \sigma\pm 2 \sigma
   _0 \right)+3 \cosh 2 \sigma \right)}{4{\cosh}\left( \sigma _0\right) {\cosh}^3\left(2 \sigma\pm\sigma_0 \right)}$}\, .\nonumber
   \end{align}
The string dynamics is governed by the type IIB Green-Schwarz action, whose bosonic part is the usual
Nambu-Goto action
\begin{gather}
  \label{bosonic_action}
 S_B = T\int d\tau d\sigma \sqrt{h}\equiv \int d\tau d\sigma \mathcal{L}_B
\end{gather}
in which $h$ is the determinant of the induced metric \eqref{induced_metric} and the string tension $T=\frac{\sqrt{\lambda}}{2\pi}$ depends on the 't~Hooft coupling $\lambda$. The leading contribution to the string partition function comes from the regularized  classical area~\cite{Drukker:2006ga} 
\begin{flalign}
  \label{classical_action}
 S_B^{(0)}(\theta_0) 
= \frac{\sqrt{\lambda}}{2\pi} \int_{0}^{2\pi} d\tau\int_{\epsilon_0}^{\infty}d\sigma\left[\sin^{2}\theta(\sigma)+\sinh^{2}\rho(\sigma)\right] 
   =\sqrt{\lambda}\left(\mp\cos\theta_{0}+\frac{1}{\epsilon}+\mathcal{O}(\epsilon)\right)~.
 \end{flalign}
 Following \cite{Kruczenski:2008zk} we have chosen to distinguish the cutoff $\epsilon_0$ in the worldsheet coordinate from the cutoff $\epsilon=\tanh\epsilon_0$ in the Poincar\'{e} radial coordinate $z$  of $\ads$.
The pole in the IR cutoff $\epsilon$   in \eqref {classical_action} keeps track of the boundary singularity of the $\ads$ metric and  it is proportional to the circumference of the boundary circle.  The standard regularization scheme, equivalent to consider a Legendre transform of the action \cite{Drukker:1999zq, Drukker:2005kx}, consists in adding a  term $-\sqrt{\lambda} \chi_b$ proportional to the boundary part of the Euler number
\begin{eqnarray} \label{Eulerboundary}
\chi_b(\theta_0)&=&\frac{1}{2\pi}\int ds  \,\, {\kappa}_g \\\nonumber
&=&\frac{3-\cosh (2 \epsilon_0)+\cosh(2\epsilon_0\pm2\sigma_0)+\cosh (4\epsilon_0\pm2\sigma_0)}{4 \sinh \epsilon_0 \cosh (\epsilon_0\pm\sigma_0) \cosh (2\epsilon_0\pm \sigma_0)}
= \frac{1}{\epsilon}\,+\mathcal{O}(\epsilon) .
\end{eqnarray}
Here $\kappa_g$ stands for the geodesic curvature of the boundary at $\sigma=\epsilon_0$ and $ds$ is the invariant line element.
With this subtraction, we have the value of the regularized classical area
\begin{equation} \label{classical_action}
  S^{\left(0\right)}_B(\theta_0)-\sqrt{\lambda} \chi_b(\theta_0)
  = \mp\sqrt{\lambda} \cos\theta_{0}~,
\end{equation}
The (upper-sign) solution dominates the string path integral and is responsible for the leading exponential behaviour in \eqref{mainratio} and so,  in the following, we will restrict to the upper signs  in \eqref{rho_and_theta}.

\section{One-loop fluctuation determinants} 
\label{sec:determinants}
This section focusses on the semiclassical expansion of the string partition function around the stable classical solution 
\eqref{background_new} (taking upper signs in \eqref{rho_and_theta}) and the determinants of the differential operators describing the semiclassical fluctuations around it. 
The  $2\pi$-periodicity in $\tau$ 
allows to trade the 2D spectral problems with infinitely-many 1D spectral problems for the (Fourier-transformed in $\tau$) differential operators in $\sigma$. Let us call $\mathcal{O}$ one of these one-loop operators. For each Fourier mode $\omega$, the evaluation of the determinant  $\textrm{Det}_{\omega}\mathcal{O}$ is a one-variable eigenvalue problem on the semi-infinite line $\sigma\in[0,\infty)$ which we solve  using the Gel'fand-Yaglom method,  a technique based on $\zeta$-function regularization reviewed in Appendix \ref{app:gelfand_yaglom}. 
Multiplying over all frequencies $\omega$ (which are integers or semi-integers according to the periodicity of the operator $\mathcal{O}$) gives then the full determinant 
\\
\begin{equation}
\label{detomega}
\textrm{Det}\mathcal{O}=\prod_{\omega}\textrm{Det}_{\omega}\mathcal{O}.
\end{equation}
All our worldsheet operators are intrinsically singular on this range of $\sigma$, since their principal symbol  diverges at $\sigma=0$, the physical singularity of the boundary  divergence for the $\ads_5$ metric. Moreover the interval is non-compact, making the spectra continuous and more difficult to deal with. We consequently introduce an IR cutoff at $\sigma=\epsilon_0$ (related to the $\epsilon=\tanh\epsilon_0$ cutoff in $z$) and one at large values of $\sigma=R$~\cite{Kruczenski:2008zk}. While the former is necessary in order to tame the near-boundary singularity, the latter has to be regarded as a mere regularization artifact descending from a small fictitious boundary on the tips of the surfaces in $\ads_3$ and $\sphere^2$. Indeed it disappears in the one-loop effective action.

\subsection{Bosonic sector}
\label{subsec:bosonicdets}

The derivation of the bosonic fluctuation Lagrangian around the minimal-area surface (\ref{background_new}) is readily available in Section 5.2 of~\cite{Forini:2015mca}. The one-loop fluctuation Lagrangian in static gauge is
\begin{gather}
\mathcal{L}^{\left(2\right)}_B\equiv\Omega^2(\sigma)\, y^T \,
  \mathcal{O}_B\left(\theta_0\right) \,y~,
\end{gather}
where the differential operator $\mathcal{O}_B\left(\theta_0\right)$ acts 
on the vector of fluctuation fields orthogonal to the worldsheet $y\equiv\left(y_i\right)_{i=1,... 
8}$. In components it reads~\footnote{To compare with~\cite{Forini:2015mca}, and using the notation used therein, notice that  the bosonic Lagrangian is derived as
\begin{eqnarray}
&& \mathcal L_B^{(2)}= \delta^{\alpha\beta}\p_\alpha y_i \p_\beta y^i - \delta^{\alpha\beta} \left(\p_\alpha y^i A_{\beta \, ij} y^j + A^i_{\alpha\, j}y^j \p_\beta y_i\right)+
\left(\delta^{\alpha\beta}A^\ell_{\alpha\, i}A_{\beta\, \ell j} -\sqrt{\gamma} \mathcal M_{i j}\right) y^i y^j \,,\quad 
\end{eqnarray}
which defines  in an obvious way $m_{ij}$ and $n_{ij}$ in \eqref{Lagrbos}.
}
\begin{gather}\label{Lagrbos}
\left[\mathcal{O}_B\left(\theta_0\right)\right]_{ij}=-\frac{1}{\Omega^2(\sigma)} \delta_{ij} 
\left(\partial_\tau^2+\partial_\sigma^2\right)+m_{ij} 
+ n_{ij} \partial_\tau\,,
\end{gather}
where the non-vanishing entries of the matrices are~\footnote{There would be an overall minus sign in the kinetic and mass term of the $y_1$ fluctuation, which we disregard in \eqref{Lagrbos} for simplifying the formula, considering that it does not play a practical role in the evaluation of determinants with Gel'fand-Yaglom and is reabsorbed in the Wick-rotation of the time coordinate $t$.}
\begin{equation}
\begin{split}
\!\!\!\!\!\!\!\!\! 
m_{11}&=m_{22}=m_{33}=  \frac{2}{\Omega^2(\sigma)\,\sinh^2\sigma}\,,\qquad
 m_{44}=  m_{55}=m_{66}=-\frac{2}{\Omega^2(\sigma)\,\cosh^2\left(\sigma+\sigma_0\right)}\,, \\
\!\!\!\!\!\!\!\!\!
 m_{77}&=  m_{88}=          \frac{-2+3\tanh^{2}\left(2\sigma+\sigma_{0}\right)}{\Omega^2(\sigma)}, \qquad\!\!\!\!\!
 n_{78} = -n_{87}=   \frac{2\tanh\left(2\sigma+\sigma_{0}\right)}{\Omega^2(\sigma)}\,.
\end{split}
\end{equation}
The worldsheet excitations decouple in the bosonic sector, apart from $y_7$ and $y_8$ which are coupled through 
 a $2\times 2$ matrix-valued differential operator.  
 The determinant of the bosonic operator is decomposed into the product 
\begin{gather}
\label{prodbos}
\textrm{Det}\mathcal{O}_B\left(\theta_0\right)=
 \textrm{Det}^3\mathcal{O}_1\,
\textrm{Det}^3\mathcal{O}_2\left(\theta_0\right)\,
\textrm{Det}\mathcal{O}_3\left(\theta_0\right).
\end{gather}
Going to Fourier space ($\partial_{\tau}\to i\omega$),  formula \eqref{prodbos}  holds for each frequency $\omega$ with
\begin{eqnarray}\label{O1}
\!\!\!
\mathcal{O}_1& \equiv& 
 -\partial^2_\sigma+\omega^2+\frac{2}{\sinh^{2}\sigma}\\\label{O2}
\!\!\!
\mathcal{O}_2\left(\theta_0\right) & \equiv&
 -\partial^2_\sigma+\omega^2-\frac{2}{\cosh^{2}\left(\sigma+\sigma_{0}\right)} \\\label{O3}
\!\!\!
\mathcal{O}_3\left(\theta_0\right) & \equiv&
 \left(\begin{array}{cc}
 -\partial^2_\sigma+\omega^2-2+3\tanh^{2}\left(2\sigma+\sigma_{0}\right) & 2\,i\,\tanh\left(2\sigma+\sigma_{0}\right)\omega\\
-2\,i\,\tanh\left(2\sigma+\sigma_{0}\right)\omega& -\frac{d^2}{d\sigma^2}+\omega^2-2+3\tanh^{2}\left(2\sigma+\sigma_{0}\right) 
\end{array}\right)
\end{eqnarray}
The unitary matrix $U=\frac{1}{\sqrt{2}}\big( \begin{smallmatrix} i & 1\\
-i & 1 \end{smallmatrix}\big)$ diagonalizes the operator \eqref{O3} 
 \begin{eqnarray}\label{O3plusminus}
 \mathcal{O}_3\left(\theta_0\right)&=&U^{\dagger}\,{\rm diag}\{ \mathcal{O}_{3+}, \mathcal{O}_{3-} \}\,U\,,\nonumber\\
 \mathcal{O}_{3+}\left(\theta_0\right)&=&-\partial_{\sigma}^{2}+\omega^{2}-2+3\tanh^{2}\left(2\sigma+\sigma_{0}\right)-2\omega\tanh\left(2\sigma+\sigma_{0}\right)\,,\\
 \mathcal{O}_{3-}\left(\theta_0\right)&=&-\partial_{\sigma}^{2}+\omega^{2}-2+3\tanh^{2}\left(2\sigma+\sigma_{0}\right)+2\omega\tanh\left(2\sigma+\sigma_{0}\right)~.\nonumber
 \end{eqnarray} 
We performed a rescaling by $\sqrt{h}=\Omega^2(\sigma)$ (as in the analogous computations of~\cite{Kruczenski:2008zk,Forini:2010ek,Drukker:2011za})  which will not affect the final determinant ratio \eqref{Zlatitude} (see discussions in Appendix A of~\cite{Drukker:2000ep} and in \cite{Kruczenski:2008zk,Forini:2010ek,Drukker:2011za}) and is actually instrumental for the analysis in Appendices~\ref{app:gelfand_yaglom_formulas} and \ref{sec:square}.

\noindent
We  rewrite \eqref{prodbos} as follows
\begin{gather}
\label{bosonic_operator_product}
\textrm{Det}_{\omega} \mathcal{O}_B\left(\theta_0\right)=
 \textrm{Det}_{\omega}^3\mathcal{O}_1\,
\textrm{Det}_{\omega}^3\mathcal{O}_2\left(\theta_0\right)\,
\textrm{Det}_{\omega}\mathcal{O}_{3+}\left(\theta_0\right)\,
\textrm{Det}_{\omega}\mathcal{O}_{3-}\left(\theta_0\right)~,
\end{gather}
where all the determinants are taken at fixed $\omega$. To reconstruct the complete bosonic contribution we have to perform an
infinite product over all possible frequencies. 

\noindent
The operator $\mathcal{O}_1$ does not depend on $\theta_0$, and indeed also appears among the 
circular Wilson loop fluctuation operators~\cite{Kruczenski:2008zk}.  While its contribution formally cancels in the ratio  \eqref{mainratiolog}, we report it below along with the others for completeness.
Both $\mathcal{O}_2\left(\theta_0\right) $ and $\mathcal{O}_3\left(\theta_0\right) $ 
become massless (scalar- and matrix-valued respectively) operators in the circular Wilson loop limit, which is clear for $\mathcal{O}_3\left(\theta_0\right)$ upon diagonalization \emph{and} an integer shift in $\omega$~\footnote{In the language of~\cite{Forini:2015mca}, this shift corresponds to a different choice of orthonormal vectors that are orthogonal to the string surface.}, irrelevant for the determinant at given frequency, as long as we do not take products over frequencies into consideration.
 Thus, in this limit one recovers the bosonic partition function of~\cite{Kruczenski:2008zk}.

 The evaluation of one-dimensional spectral problems is outlined in Appendix \ref{app:gelfand_yaglom_formulas}. The fields satisfy Dirichlet boundary conditions at the endpoints of the compactified interval 
$\sigma\in[\epsilon_0, R]$. Then we take the limit of the value of the regularized determinants for $R\to\infty$ at fixed $\omega$ and $\epsilon_0$. As evident from the expressions below, the limit on the physical  IR cutoff ($\epsilon$ in $z$ or equivalently $\epsilon_0$ in $\sigma$) would drastically change the $\omega$-dependence at this stage and thus would spoil the product over the frequencies.  It is a crucial, a posteriori, observation that it is only keeping $\epsilon_0$ \emph{finite} while sending $R$ to infinity that one precisely reproduces the expected large $\omega$ (UV) divergences~\cite{Drukker:2000ep, Forini:2015mca}. This comes at the price of more complicated results for the bosonic (and especially fermionic) determinants. Afterwards we will remove the IR divergence in the one-loop effective action by referring the latitude to the circular solution.

The solutions of the differential equations governing the different determinants are singular  for small subset of frequencies
: We  shall  treat apart  these special values when reporting the solutions. For the determinant of the operator $\mathcal{O}_1$ in \eqref{O1} in the limit of large $R$ one obtains~\cite{Kruczenski:2008zk}
\begin{flalign}\label{detO1}
\textrm{Det}_{\omega} \mathcal{O}_1=\begin{cases}
e^{|\omega|  (R-\epsilon_0 )} \frac{(|\omega| +\coth\epsilon_0)}{2 |\omega|  (|\omega| +1)}& 
\,\qquad\qquad \omega\neq0\\
R \coth\epsilon_0&
\,\qquad\qquad \omega=0
\end{cases}
\end{flalign}
and only the case $\omega=0$ has to be considered separately. Next we examine 
the initial value problem (\ref{O2first})-(\ref{O2second}) associated to $\mathcal{O}_2(\theta_0)$, whose solution is 
\begin{flalign}
f_{(II)1}(\sigma) =\begin{cases}
\frac{1}{2 \omega  \cosh \left(\sigma +\sigma_0\right) \cosh \left(\sigma_0+\epsilon_0\right)}
\left(\cosh \left(\sigma+\epsilon_0 +2 \sigma_0 \right) \sinh (\omega  (\sigma -\epsilon_0 )) 
\vphantom{\frac{(\omega -1) \sinh ((\omega +1) (\sigma -\epsilon_0 ))}{2 (\omega +1)}}+
\right.&\\
\qquad\qquad\qquad \left.+\frac{(\omega +1) \sinh ((\omega -1) (\sigma -\epsilon_0 ))}{2 (\omega -1)}+\frac{(\omega -1) \sinh ((\omega +1) (\sigma -\epsilon_0 ))}{2 (\omega +1)}\right) & 
\,\  \omega\neq-1,0,1\\
\frac{2\left(\sigma-\epsilon_0\right)-\sinh2\left(\sigma_{0}+\epsilon_0\right)+\sinh2\left(\sigma+\sigma_{0}\right)}{4\cosh\left(\sigma_{0}+\epsilon_0\right)\cosh\left(\sigma+\sigma_{0}\right)}&
\,\   \omega=-1,1\\
\frac{\left(\sigma-\epsilon_0\right)\sinh\left(\sigma_{0}+\epsilon_0\right)\sinh\left(\sigma+\sigma_{0}\right)+\sinh\left(\sigma-\epsilon_0\right)}{\cosh\left(\sigma_{0}+\epsilon_0\right)\cosh\left(\sigma+\sigma_{0}\right)}&
\,\  \omega=0 \,.
\end{cases}
\end{flalign}
The determinant is then given by $f_{(II)1}(R)$ and for $R$ large one obtains the simpler expression 
\begin{flalign}\label{detO2}
\textrm{Det}_{\omega} \mathcal{O}_2(\theta_0)=\begin{cases}
e^{|\omega|  (R-\epsilon_0 )}\frac{ (|\omega| +\tanh (\sigma_0+\epsilon_0 ))}{2 |\omega|  (|\omega| +1)}\,& 
\,\qquad\qquad \omega\neq0\\
R\,\tanh(\sigma_0+\epsilon_0) &
\,\qquad\qquad \omega=0\,.
\end{cases}
\end{flalign}
We repeat the same procedure for $\mathcal{O}_{3+}\left(\theta_0\right)$. From the solutions
\begin{flalign}
f_{(II)1}(\sigma) =\begin{cases}
\frac{(\omega +1) e^{2 \left(\sigma +\sigma_0+\epsilon_0\right)} \sinh \left((\omega -1) \left(\sigma-\epsilon_0 \right)\right)+(\omega -1) \sinh \left((\omega +1) \left(\sigma-\epsilon_0 \right)\right)}
   {\left(\omega ^2-1\right)\sqrt{(1+e^{4 \sigma+2 \sigma_0})(1+e^{2 \sigma_0+4 \epsilon_0})}} & 
\,\qquad \omega\neq-1,0,1\\
\frac{\left(e^{2 \sigma -2\epsilon_0}-e^{-2\sigma+2 \epsilon_0}+4 e^{2 \sigma +2 \sigma_0+2 \epsilon_0} \left(\sigma-\epsilon_0 \right)\right)}
   {4 \sqrt{(1+e^{4 \sigma +2 \sigma_0})(1+e^{2 \sigma_0+4 \epsilon_0})}}&
\,\qquad \omega=-1,1\\
\frac{\left(e^{ \sigma-\epsilon_0 }-e^{-\sigma+ \epsilon_0}\right) \left(e^{2
   \left(\sigma +\sigma_0+\epsilon_0\right)}+1\right)}{2 \sqrt{(1+e^{4 \sigma +2 \sigma_0})(1+e^{2 \sigma_0+4 \epsilon_0})}}
&
\,\qquad \omega=0
\end{cases}
\end{flalign}
one finds  for large $R$
\begin{flalign}\label{detO3plus}
~~
\textrm{Det}_{\omega} \mathcal{O}_{3+}(\theta_0)=\begin{cases}
\frac{e^{R (\omega-1)-\sigma_0-(\omega +1) \epsilon_0} \left(\omega +(\omega +1) e^{2 \sigma_0+4 \epsilon_0}-1\right)}
   {2 \left(\omega ^2-1\right) \sqrt{1+e^{2 \sigma_0+4 \epsilon_0}}}& 
\,\qquad\qquad \omega\geq2\\
 \frac{R e^{\sigma_0+2 \epsilon_0 }}{\sqrt{1+e^{2 \sigma_0+4 \epsilon_0 }}} &
\,\qquad\qquad \omega=1\\
 \frac{e^{-R (\omega-1 )+\sigma_0+(\omega +1) \epsilon_0}}{2(1-\omega ) \sqrt{1+e^{2 \sigma_0+4 \epsilon_0}}}&
\,\qquad\qquad \omega\leq0\,.
\end{cases}
\end{flalign}
In view of the relation  $\mathcal{O}_{3-}\left(\theta_0\right)=\mathcal{O}_{3+}\left(\theta_0\right)\mid_{\omega\to-\omega}$, which follows from \eqref{O3plusminus}, we can easily deduce the results for $\textrm{Det}_{\omega} \mathcal{O}_{3-}(\theta_0)$ by flipping the frequency  in the lines above
\begin{flalign}\label{detO3minus}
\qquad
\textrm{Det}_{\omega} \mathcal{O}_{3-}(\theta_0)=\begin{cases}
 \frac{e^{R (\omega+1 )+\sigma_0+(-\omega +1) \epsilon_0}}{2(1+\omega ) \sqrt{1+e^{2 \sigma_0+4 \epsilon_0}}}&
\,\qquad\omega\geq0\\
 \frac{R e^{\sigma_0+2 \epsilon_0 }}{\sqrt{1+e^{2 \sigma_0+4 \epsilon_0 }}} &
\,\qquad \omega=-1\\
\frac{e^{-R (\omega
   +1)-\sigma_0-(-\omega +1) \epsilon_0} \left(-\omega +(-\omega +1) e^{2 \sigma_0+4 \epsilon_0}-1\right)}{2 \left(\omega ^2-1\right) \sqrt{1+e^{2 \sigma_0+4 \epsilon_0}}}& 
\,\qquad  \omega\leq-2\,.
\end{cases}
\end{flalign}
Notice that a shift of $\omega\to\omega-1$ in $\textrm{Det}_\omega\mathcal{O}_{3+}(\theta_0)$ and $\omega\to\omega+1$ in $\textrm{Det}_\omega\mathcal{O}_{3-}(\theta_0)$ gives back the  symmetry around $\omega=0$ 
in the distribution of power-like and exponential large-$R$ divergences which characterizes the other determinants 
\eqref{detO1} and \eqref{detO2}. Such a shift -- also useful for the circular Wilson loop limit as discussed below \eqref{bosonic_operator_product} -- does not affect the 
determinant, and we will perform it in Section \ref{sec:partitionfunctions}.
 
\bigskip


\subsection{Fermionic sector}
\label{sec:fermionic_Lagrangian}

The fluctuation analysis in the fermionic sector can be easily carried out following again the general approach~\cite{Forini:2015mca}, which includes the local $\grSO(1,9)$ rotation in the target space~\cite{Kavalov:1986nx,Sedrakian:1986np,Langouche:1987mw,Langouche:1987my,Langouche:1987mx,Drukker:2000ep} that allows to cast the quadratic Green-Schwarz fermionic action into eight contributions for two-dimensional spinors on the curved worldsheet background. 

The standard Type IIB $\kappa$-symmetry gauge-fixing for the rotated fermions $\Psi^I$
\begin{equation}
  \label{kappa_symmetry}
\Psi^1=\Psi^2\equiv\Psi\,
\end{equation}
leads to the Lagrangian~\footnote{We perform the computations in a Lorentzian signature for the induced worldsheet metric and only at the end Wick-rotate back. The difference with (5.37)-(5.38) of~\cite{Forini:2015mca} is only in labeling the spacetime directions.}
 \begin{eqnarray} 
  \mathcal{L}_F^{\left(2\right)} &=& 2i \,\Omega^2(\sigma)\,\bar{\Psi}\, \mathcal{O}_F\left(\theta_0\right)\,\Psi\,
  \end{eqnarray}
  where the operator  $\mathcal{O}_F\left(\theta_0\right)$ is given by
  \begin{eqnarray}
   \label{fermionic_operator_initial}
 \mathcal{O}_F\left(\theta_0\right) &=&\frac{i}{\Omega(\sigma)}\left(\Gamma_{4}\partial_{\tau}+\Gamma_{3}\partial_{\sigma}-a_{34}(\sigma)\Gamma_{3}+a_{56}(\sigma)\Gamma_{456}\right)\nonumber\\
&&+\frac{1}{\Omega (\sigma)^{2}}\left(\sinh^{2}\rho(\sigma)\Gamma_{012}+\sin^{2}\theta(\sigma)\Gamma_{0123456}\right).\, \end{eqnarray}
The coefficients   $a_{34} (\sigma )$ and $a_{56} (\sigma )$  can be expressed as derivatives of the  functions appearing in the classical solution:
\begin{flalign}
a_{34} (\sigma ) & 
=-\frac{1}{2}\frac{d}{d\sigma}\log\Omega (\sigma )\quad\mathrm{and}\quad
a_{56}(\sigma)  
=\frac{1}{4}\frac{d}{d\sigma}\log\frac{\cosh\rho(\sigma)+\cos\theta(\sigma)}{\cosh\rho(\sigma)-\cos\theta(\sigma)}.
\end{flalign}
In the $\theta_0\to0$ limit (hence $\theta(\sigma)\to0$), one gets
\begin{eqnarray}
\mathcal{O}_{F}\left(\theta_0=0\right)
  \label{fermionic_action_circle_us}
=i\sinh\sigma\Gamma_{4}\partial_{\tau}+i\sinh\sigma\Gamma_{3}\partial_{\sigma}-\frac{i}{2}\cosh\sigma\Gamma_{3}+\frac{i}{2}\sinh\sigma\Gamma_{456}+\Gamma_{012}\,,
\end{eqnarray}
which coincides with the operator 
found in the circular Wilson loop analysis of~\cite{Kruczenski:2008zk}~\footnote{See formula (5.17) therein.},
once we go back to Minkowski signature and  reabsorbe the connection-related 
$\Gamma_{456}$-term via the $\tau$-dependent rotation~$\Psi\to\exp\left(-\frac{\tau}{2}\Gamma_{56}\right)\Psi$. 
In Fourier space 
this results in a shift of the integer fermionic frequencies $\omega$ by one half, 
turning periodic fermions into anti-periodic ones. 
In the general case \eqref{fermionic_operator_initial}
we cannot eliminate all the connection-related terms
$-a_{34}(\sigma)\Gamma_{3}+a_{56}(\sigma)\Gamma_{456}$, 
since the associated normal bundle is non-flat~\cite{Forini:2015mca}~\footnote{The arising 
of gauge connections in the covariant derivatives associated to the structure of normal 
bundle is discussed at length in~\cite{Forini:2015mca} and references therein. 
In particular, see discussion in Section 5.2 of~\cite{Forini:2015mca} for both the latitude and 
the circular Wilson loop limit.}. Performing anyway the above $\tau$-rotation  
at the level of \eqref{fermionic_operator_initial} has  the 
merit of simplifying the circular limit making a direct connection with known results.
This is how we will proceed: For now, we  continue with the analysis of the fermionic operator in the form \eqref{fermionic_operator_initial} without performing any rotation. Then,
in Section \ref{sec:partitionfunctions}, we  shall take into account the effect of this rotation by relabelling 
the fermionic Fourier modes   in terms of a suitable choice of half-integers. 

The analysis of the fermionic operator \eqref{fermionic_operator_initial} drastically simplifies
noticing that the set of mutually-commuting matrices $\{\Gamma_{12},\Gamma_{56},\Gamma_{89}\}$ commutes with the operator itself and leaves invariant the spinor constraint \eqref{Weyl} and the fermionic gauge fixing (\ref{kappa_symmetry}).  By means of the projectors
\begin{flalign}
\mathcal{P}_{12}^{\pm}  \equiv\frac{\mathbb{I}_{32}\pm i\Gamma_{12}}{2},\ \ \ \ 
\mathcal{P}_{56}^{\pm}  \equiv\frac{\mathbb{I}_{32}\pm i\Gamma_{56}}{2}\ \ \ \ \mathrm{and}\ \ \  \
\mathcal{P}_{89}^{\pm}  \equiv\frac{\mathbb{I}_{32}\pm i\Gamma_{89}}{2},
\end{flalign}
we decompose the $32\times 32$ fermionic operator into eight blocks of $2\times 2$ operators labeled by the triplet $\{p_{12},p_{56},p_{89}=-1,1\}$. 
Formally this can be seen as the decomposition into  the following orthogonal subspaces 
\begin{flalign}
\mathcal{O}_{F}\left(\theta_0\right)&=
\underset{p_{12},p_{56},p_{89}=-1,1}{\bigoplus}\mathcal{O}_{F}^{p_{12},p_{56},p_{89}}\left(\theta_0\right)\\
\Psi&=
\underset{p_{12},p_{56},p_{89}=-1,1}{\bigotimes}\Psi^{p_{12},p_{56},p_{89}}
\end{flalign}
where each operator
\begin{eqnarray}\label{fermionic_operator_decomposed}
\mathcal{O}_{F}^{p_{12},p_{56},p_{89}}\left(\theta_0\right)
&\equiv&
\frac{i}{\Omega(\sigma)}\left(\Gamma_{4}\partial_{\tau}+\Gamma_{3}\partial_{\sigma}-a_{34}(\sigma)\Gamma_{3}-ip_{56}a_{56}(\sigma)\Gamma_{4}\right)\\
&&+\frac{1}{\Omega^{2}(\sigma)}\left(-ip_{12}\sinh^{2}\rho(\sigma)\Gamma_{0}-p_{12}p_{56}\sin^{2}\theta(\sigma)\Gamma_{034}\right)\nonumber
\end{eqnarray}
acts on the eigenstates $\Psi^{p_{12},p_{56},p_{89}}$ of $\{\mathcal{P}_{12}^{\pm},\mathcal{P}_{56}^{\pm},\mathcal{P}_{89}^{\pm}\}$ with
eigenvalues $\{\frac{1\pm \,p_{12}}{2},\frac{1\pm \,p_{56}}{2},\frac{1\pm \,p_{89}}{2}\}$. 
Notice that the operator defined in \eqref{fermionic_operator_decomposed} actually does not depend on the label $p_{89}$. 
Then the spectral problem reduces to the computation of eight 2D functional determinants \footnote{A non-trivial matrix structure is also encountered in the fermionic sector of the circular Wilson loop~\cite{Kruczenski:2008zk}, but the absence of a background geometry in $\sphere^5$ leads to a simpler gamma structure. It comprised only three gamma combinations ($\Gamma_0,\Gamma_4,\Gamma_{04}$), whose algebra allows their identification with the three Pauli 
matrices without the need of the labelling the subspaces.}
\begin{gather}
\label{prod_fermionic_operator_decomposed}
\textrm{Det}\mathcal{O}_{F}\left(\theta_0\right)
=\prod_{p_{12},p_{56},p_{89}=\pm1}\textrm{Det}\mathcal{O}_{F}^{p_{12},p_{56},p_{89}}\left(\theta_0\right).
\end{gather}
A deeper look at   the properties of  $\mathcal{O}_{F}^{p_{12},p_{56},p_{89}}$  allows us to  focus just on the case  
of $p_{12}=p_{56}=p_{89}=1$. In fact, as motivated in details  in Appendix \ref{ferm111},  the total determinant can be rewritten as follows 
\be\label{detfermall1mirror}
\textrm{Det}\mathcal{O}_{F}\left(\theta_0\right)=\prod_{\omega\in\mathbb{Z}}\,\textrm{Det}_{\omega}[(\mathcal{O}_{F}^{{1},1,1}(\omega) )^2]^2 
\textrm{Det}_{\omega}[(\mathcal{O}_{F}^{{1},1,1}(-\omega) )^2]^2~\,.
\ee
Using the matrix representation (\ref{representation_gamma}) and going to Fourier space, we obtain 
\begin{eqnarray}\label{O111}
\mathcal{O}_{F}^{1,1,1}(\theta_0)
&\equiv&\Big[
\frac{i}{\Omega(\sigma)}\big(-i\omega \sigma_2+\sigma_{1}\partial_{\sigma}-a_{34}(\sigma)\sigma_{1}+ia_{56}(\sigma)\sigma_2\big)\\\nonumber
&&+\frac{1}{\Omega^{2}(\sigma)}\big(\sinh^{2}\rho(\sigma)\sigma_{3}-\sin^{2}\theta(\sigma){\mathbb{I}_2}\big)\Big]\otimes M
\equiv  \mathcal{\widetilde{O}}_{F}^{1,1,1}\otimes M~,
\end{eqnarray}
where $M=\sigma_2\otimes {\mathbb{I}_4}\otimes\sigma_1$. 
For simplicity of notation, from now on we  will  denote with $O_{F}^{1,1,1}(\theta_0)$ the first factor in the definition above.
In a similar spirit to the analysis for the bosonic sector, we start to find the solutions of the homogeneous problem
\begin{gather}
  \label{homogeneous_equation_fermions}
 \mathcal{{O}}_{F}^{1,1,1}\left(\theta_0\right) 
\bar{f}(\sigma)=0
\end{gather}
where
$\bar{f}(\sigma)$  denotes the two component spinor $ \left(f_{1}(\sigma), f_{2}(\sigma)\right)^T$. The system of coupled first-order differential equations  now reads
\begin{eqnarray}
  \label{first_equation}
\!\!\!\!\!\!\!\!\!\!\!\!\!\!\!\!\!\!\!\!\!\!\!\!\!
&&\left(- \sin^{2}\theta(\sigma)+\sinh^{2}\rho(\sigma)\right)f_{1}(\sigma)+i \Omega(\sigma)\left(\partial_{\sigma}-\omega-a_{34}(\sigma)+a_{56}(\sigma)\right)f_{2}(\sigma)=0,\\
  \label{second_equation}
\!\!\!\!\!\!\!\!\!\!\!\!\!\!\!\!\!\!\!\!\!\!\!\!\!
&&\left(-\sin^{2}\theta(\sigma)-\sinh^{2}\rho(\sigma)\right)f_{2}(\sigma)+i \Omega(\sigma)\left(\partial_{\sigma}+\omega-a_{34}(\sigma)-a_{56}(\sigma)\right)f_{1}(\sigma)=0.
\end{eqnarray}
We can cast it into a second-order differential equation for one of the unknown functions.  
Solving  (\ref{second_equation}) for $f_2(\sigma)$ 
\begin{gather}
f_{2} (\sigma)=\frac{i }{\Omega\left(\sigma\right)}\left(\partial_{\sigma}+\omega-\frac{1}{2\tanh\sigma}-\frac{\tanh\left(\sigma+\sigma_{0}\right)}{2}\right)f_{1}(\sigma)\,,
\end{gather}
and then plugging it into (\ref{first_equation}) one obtains
\begin{gather}
    \label{Schoedinger_p56PLUS}
\!\!\!
 f_{1}^{''}(\sigma)-\Big(\frac{1}{2\sinh^{2}\sigma}-\frac{1}{2\cosh^{2}\left(\sigma+\sigma_{0}\right)}+\Big(\frac{1}{2\tanh\sigma}+\frac{\tanh(\sigma+\sigma_{0})}{2}-\omega\Big)^{2}\Big)\,f_{1}(\sigma)=0.
\end{gather}
It is worth noticing that the Gel'fand-Yaglom method has naturally led to an auxiliary Schr\"{o}dinger equation  for a fictitious 
particle on a semi-infinite line and subject to a supersymmetric potential $V(\sigma)=-W'(\sigma)+W^2(\sigma)$
derived from the prepotential $W(\sigma)=\frac{1}{2\tanh\sigma}+\frac{\tanh(\sigma+\sigma_{0})}{2}-\omega$.
Traces of supersymmetry are not surprising: They represent  a vestige of the supercharges unbroken by the 
classical background~\footnote{The same property is showed by (5.26) in 
\cite{Kruczenski:2008zk}. }.\\
\\
As in the bosonic case, we have to separately discuss some critical values of the frequencies. We only report the independent solutions of the equations above, where the constants $c_{i,1}$ and $c_{i,2}$ have to be fixed in the desired initial value problem ($i=I,\,II$). 
\begin{flalign} \label{ferm_sol1}
f_{(i)1}(\sigma) =\begin{cases}
\frac{c_{i,1}e^{\sigma\left(1+\omega\right)}+c_{i,2}e^{\sigma\left(1-\omega\right)+\sigma_{0}}\left(2\omega^{2}\cosh\left(\sigma+\sigma_{0}\right)\sinh\sigma+\omega\cosh\left(2\sigma+\sigma_{0}\right)+\sinh\sigma_{0}\right)}{\sqrt{\left(e^{2\sigma}-1\right)\left(e^{2\sigma+2\sigma_{0}}+1\right)}} &
\,\omega\neq-1,0,1\\
\frac{c_{i,1}e^{2\sigma}+c_{i,2}\left(-4\sigma e^{2\sigma+2\sigma_{0}}-2e^{2\sigma_{0}}+2-e^{-2\sigma}\right)}{\sqrt{\left(e^{2\sigma}-1\right)\left(e^{2\sigma+2\sigma_{0}}+1\right)}} &
\ \omega=1\\
\frac{c_{i,1}e^{\sigma}+c_{i,2}\left(-e^{-\sigma}-e^{3\sigma+2\sigma_{0}}+2\sigma e^{\sigma}\left(e^{2\sigma_{0}}-1\right)\right)}{\sqrt{\left(e^{2\sigma}-1\right)\left(e^{2\sigma+2\sigma_{0}}+1\right)}} &
\, \omega=0\\
\frac{c_{i,1}+c_{i,2}\left(4\sigma-2e^{2\sigma}+2e^{2\sigma+2\sigma_{0}}-e^{4\sigma+2\sigma_{0}}\right)}{\sqrt{\left(e^{2\sigma}-1\right)\left(e^{2\sigma+2\sigma_{0}}+1\right)}} &
\, \omega=-1\\
\end{cases}
\end{flalign}
\begin{flalign} \label{ferm_sol2}
f_{(i)2}(\sigma) =\begin{cases}
c_{i,1}\frac{2i e^{\sigma\left(2+\omega\right)+\sigma_{0}}\left(-\cosh\left(2\sigma+\sigma_{0}\right)+2\omega\cosh\left(\sigma+\sigma_{0}\right)\sinh\sigma\right)}{\sqrt{\left(e^{2\sigma}-1\right)\left(e^{2\sigma_{0}}+1\right)\left(e^{2\sigma+2\sigma_{0}}+1\right)\left(e^{4\sigma+2\sigma_{0}}+1\right)}}+ &\\
\qquad  -c_{i,2}\frac{i e^{\sigma\left(2-\omega\right)+2\sigma_{0}}\left(2\omega+\sinh\left(2\sigma+2\sigma_{0}\right)+2\omega\sinh\sigma_{0}\sinh\left(2\sigma+\sigma_{0}\right)-\sinh2\sigma\right)}{\sqrt{\left(e^{2\sigma}-1\right)\left(e^{2\sigma_{0}}+1\right)\left(e^{2\sigma+2\sigma_{0}}+1\right)\left(e^{4\sigma+2\sigma_{0}}+1\right)}}&
\!\!\!\!\!\!\!\!\!\!\!\!\omega\neq-1,0,1\\
-c_{i,1}\frac{i  \left(2e^{\sigma}-e^{3\sigma}+e^{3\sigma+2\sigma_{0}}\right)}{\sqrt{\left(e^{2\sigma}-1\right)\left(e^{2\sigma_{0}}+1\right)\left(e^{2\sigma+2\sigma_{0}}+1\right)\left(e^{4\sigma+2\sigma_{0}}+1\right)}}  &\\
\qquad -c_{i,2}\frac{i  \left(2e^{5\sigma+4\sigma_{0}}-e^{-\sigma+2\sigma_{0}}+e^{-\sigma}+4\left(\sigma+1\right)e^{3\sigma+2\sigma_{0}}\left(1-e^{2\sigma_{0}}\right)-4\left(2\sigma+1\right)e^{\sigma+2\sigma_{0}}\right)}{\sqrt{\left(e^{2\sigma}-1\right)\left(e^{2\sigma_{0}}+1\right)\left(e^{2\sigma+2\sigma_{0}}+1\right)\left(e^{4\sigma+2\sigma_{0}}+1\right)}} &
\, \omega=1\\
-c_{i,1}\frac{i  \sqrt{1+e^{4\sigma+2\sigma_{0}}}}{\sqrt{\left(e^{2\sigma}-1\right)\left(e^{2\sigma_{0}}+1\right)\left(e^{2\sigma+2\sigma_{0}}+1\right)}} &\\
\qquad  -c_{i,2}\frac{i  \left(e^{2\sigma}-6e^{2\sigma+2\sigma_{0}}+e^{2\sigma+4\sigma_{0}}+2\left(\sigma-1\right)e^{4\sigma+2\sigma_{0}}\left(e^{2\sigma_{0}}-1\right)+2\left(\sigma+1\right)\left(e^{2\sigma_{0}}-1\right)\right)}{\sqrt{\left(e^{2\sigma}-1\right)\left(e^{2\sigma_{0}}+1\right)\left(e^{2\sigma+2\sigma_{0}}+1\right)\left(e^{4\sigma+2\sigma_{0}}+1\right)}}
&
\ \omega=0\\
-c_{i,1}\frac{i  \left(e^{\sigma}-e^{\sigma+2\sigma_{0}}+2e^{3\sigma+2\sigma_{0}}\right)}{\sqrt{\left(e^{2\sigma}-1\right)\left(e^{2\sigma_{0}}+1\right)\left(e^{2\sigma+2\sigma_{0}}+1\right)\left(e^{4\sigma+2\sigma_{0}}+1\right)}} &\\
\qquad -c_{i,2}\frac{i \left(2e^{-\sigma}-e^{5\sigma+2\sigma_{0}}+e^{5\sigma+4\sigma_{0}}+4\left(\sigma-1\right)e^{\sigma}\left(1-e^{2\sigma_{0}}\right)+4e^{3\sigma+2\sigma_{0}}\left(2\sigma-1\right)\right)}{\sqrt{\left(e^{2\sigma}-1\right)\left(e^{2\sigma_{0}}+1\right)\left(e^{2\sigma+2\sigma_{0}}+1\right)\left(e^{4\sigma+2\sigma_{0}}+1\right)}}&
\, \omega=-1
\end{cases}
\end{flalign}
We are now ready to evaluate the determinants using the results of Appendix \eqref{sec:square}, namely 
considering   Dirichlet boundary conditions for the square of the first order differential operator.  
Having in mind  the solutions above and how they enter in \eqref{GY_Y} and 
\eqref{OY}, it is clear that already  the \emph{integrand} in  \eqref{op_square_Dirichlet} is significantly 
complicated. A simplification occurs by recalling that our final goal is 
taking the $R\to \infty$  limit of all determinants and combine them in the ratio 
of bosonic and fermionic contributions. As stated above in the bosonic analysis 
and shown explicitly below, for the correct large $\omega$ divergences to be reproduced, it is crucial to send $R\to\infty$ while keeping
 $\epsilon$ finite. In Appendix \eqref{largeRsimplified} we sketch how to 
use the main structure of the matrix of the solutions $Y(\sigma)$ to obtain the 
desired large-$R$ expressions for the determinants in a more direct way.\\
\\
The determinant of the operator $O_F^{1,1,1}$ for modes  $\omega\neq\{-1,0,1\}$ reads for large $R$
\begin{align}\label{finalferm} 
&
\textrm{Det}_{\omega\geq2}[(\mathcal{O}_{F}^{{1},1,1})^2]~ = \frac{ a_0\,e^{2
   \omega  (R-\epsilon_0 )}}{ \omega ^2\,
   (1+\omega)^2 (\omega-1 )}\,\Big[a_1\,\Phi \left(e^{-2 \epsilon_0 },1,\omega \right)+a_2\,\Phi  (-e^{-2 (\sigma_0+\epsilon_0
   )},1,\omega )+a_3\Big]\nonumber \\
&
\textrm{Det}_{\omega\leq-2}[(\mathcal{O}_{F}^{{1},1,1})^2]=\frac{b_0\,e^{-2
   \omega  (R-\epsilon_0 )}}{ \omega \, (1-\omega )^2}\,\Big[b_1\,\Phi \left(e^{-2 \epsilon_0 },1,-\omega \right)+b_2\,\Phi  (-e^{-2 (\sigma_0+\epsilon_0
   )},1,-\omega )+b_3\Big]
\end{align}
where $\Phi(z,s,a)$ is the Lerch transcendent \eqref{lerch_def}.  The presence of the Lerch function is just 
a tool to have  a compact expression for the determinants. In fact, for the values of  $\omega$ relevant for us, it can be 
can be written  in terms of elementary functions, but its expression becomes more and more unhandy as the value of $\omega$ increases.
The coefficients $a_i$ and $b_i$ can be also expressed in terms of elementary functions. 
For the $a_i$ we have 
\begin{align}\nonumber
a_0&=e^{-R-\frac{3 \sigma _0}{2}} \,\frac{\sinh \epsilon_0  \left(\tanh\sigma _0+1\right)\cosh \left(\sigma _0+\epsilon_0
   \right)}{8  \sqrt{2\cosh
   \left(\sigma _0+2 \epsilon_0 \right)}}\,\\
a_1&=4 \,\text{sech}\sigma_0 (\tanh
   \sigma_0+\omega )^2\,\\\nonumber
a_2&=4 [2
   \left(1-\omega ^2\right) \omega ^2 \cosh
   \sigma_0-2 \left(1-\omega ^2\right) \omega 
   \sinh\sigma _0+\text{sech}\sigma _0
   \left(\text{sech}^2\sigma_0+\omega ^2-1\right)]\,\\\nonumber
  a_3 &=\tanh ^2 \sigma _0\,  (\coth\epsilon_0+1) \,\text{csch}\epsilon_0\,
   \text{sech}\,\left(\sigma _0+\epsilon_0 \right)\,
 \big[e^{\sigma _0}
   \left(\cosh   \sigma _0 -2 \sinh \sigma _0-\sinh (2
   \epsilon_0 -\sigma _0)\right)\\\nonumber 
   &+\cosh(2\sigma
   _0+2\epsilon_0))\big]+2 \omega \, \Big[-  \omega ^2 \cosh ^2 \sigma _0  \,  \text{csch}\epsilon_0\,   \text{sech} (\sigma _0+\epsilon_0)\\ \nonumber
&+   \cosh \sigma _0 \big(2 \omega ^2+\omega +3 \omega ^2\,   \text{csch}\epsilon_0\,\cosh (\sigma_0+2 \epsilon_0) \,  \text{sech}(\sigma _0+\epsilon_0)
   +\omega  \coth ^2\epsilon_0+2 \coth\epsilon_0-2\big)\\\nonumber
&
+2 \,\big(3 \omega \,\cosh \epsilon_0\,\text{sech} (\sigma _0+\epsilon_0 )   -\sinh\sigma_0\,\left(\omega -2 \omega  \coth \epsilon_0-\text{csch}^2\epsilon_0\right)-\text{sech}\sigma _0   (\coth \epsilon_0+1)\big) \Big]\, \nonumber
\end{align}
while for the $b_i$  we get
\begin{equation}
\begin{split}
b_0&=e^{R-\frac{\sigma _0}{2}}\text{sech}^2\sigma _0\, \frac{   \sinh\epsilon_0\,(\tanh\sigma _0+1)
  \cosh \left(\sigma _0+\epsilon_0
   \right)}{8 \sqrt{2\cosh(\sigma _0+2 \epsilon_0)}}\,\label{finalfermomeganotspecial} \\ 
   b_1&=-2\,\\ 
b_2&=-2 \,\big[\,\omega  \big(\omega  \cosh (2\sigma _0)+\sinh(2 \sigma_0)\big)+\omega ^2-1\,\big]\,\\
b_3&=-\cosh ^2\sigma_0\,\big[4 \omega  \tanh(\sigma _0+\epsilon_0)
-2 \omega\,\coth\epsilon_0+\text{csch}^2\epsilon_0\,\big]-\omega\\
&- \cosh(2 \sigma_0)(\omega +1)-\sinh (2 \sigma_0)+\cosh(\epsilon_0 -\sigma _0) \text{sech}\,(\sigma_0+\epsilon_0)\,.
\end{split}
\end{equation}
The determinants of the lower modes  have to be computed separately and they are given by
\begin{align}\nonumber
&
\textrm{Det}_{\omega=1}[(\mathcal{O}_{F}^{{1},1,1})^2]\!=\!
  R\,e^{R}\,\frac{e^{-\frac{\sigma_0}{2}} (\tanh\sigma_0+1) \sinh \epsilon_0 \cosh (\sigma_0+\epsilon_0)}{\left(e^{2 \sigma_0}+1\right)^3 \sqrt{2 \cosh (\sigma_0+2 \epsilon_0)}}
  \Big[-2 e^{4\sigma_0} \Big(\log \frac{e^{2 \epsilon_0 }-1}{e^{2 (\sigma_0+\epsilon_0 
)}+1}+
\\ \label{specialmode1}
&+2 \sigma_0\Big)
+\frac{(e^{2 \sigma _0}+1) \big(e^{6 \sigma _0+4 \epsilon_0 }+(e^{2 \epsilon_0 }+1) e^{4 \sigma _0+2 \epsilon_0 }
  +e^{2 \sigma _0}  (-5 e^{2 \epsilon_0 }+3 e^{4 \epsilon_0 }+3 )+ (e^{2 \epsilon_0 }-1)^2\big)}
  {(e^{2 \epsilon_0 }-1)^2  (e^{2  (\sigma _0+\epsilon_0)}+1)}
\Big]\, 
 \\\nonumber \label{specialmode0}
&
\textrm{Det}_{\omega=0}[(\mathcal{O}_{F}^{{1},1,1})^2]\!=\!
  R\,e^{R}\,\frac{e^{-\frac{\sigma_0}{2}} (\tanh\sigma_0+1) \sinh \epsilon_0 \cosh (\sigma_0+\epsilon_0)}{\left(e^{2 \sigma_0}+1\right)^2 \sqrt{2 \cosh (\sigma_0+2 \epsilon_0)}}\Big[-2 e^{2 \sigma_0} \Big(\log\frac{e^{2 \epsilon_0 }-1}{e^{2 (\sigma_0+\epsilon_0 )}+1}+\\
&+2 \sigma_0\Big)
+\frac{\left(e^{2\sigma_0}+1\right) \left(-e^{2\sigma_0}+3 e^{2 (\sigma_0+\epsilon_0 )}
+e^{4 (\sigma_0+\epsilon_0 )}-e^{2 \epsilon_0 }+e^{4 \epsilon_0 }+1\right)}{\left(e^{2 \epsilon_0 }-1\right)^2 
\left(e^{2 (\sigma_0+\epsilon_0 )}+1\right)}
\Big]\, 
\\\nonumber
&
\textrm{Det}_{\omega=-1}[(\mathcal{O}_{F}^{{1},1,1})^2]\!=\!
e^{3R}\,\frac{e^{-\frac{\sigma_0}{2}} (\tanh\sigma_0+1) \sinh \epsilon_0 \cosh (\sigma_0+\epsilon_0)}{8\,\left(e^{2 \sigma_0}+1\right)^2 \sqrt{2 \cosh (\sigma_0+2 \epsilon_0)}} \Big[-2 e^{2\sigma_0} \Big(\log \frac{e^{2 \epsilon_0 }-1}{e^{2 (\sigma_0+\epsilon_0 )}+1}
+\\\label{specialmodes}
&+2 \sigma_0\Big) 
+\frac{\left(e^{2\sigma_0}+1\right) \left(e^{4\sigma_0} \left(2 e^{2 \epsilon_0 }-1\right)+e^{2\sigma_0} \left(7 e^{2 \epsilon_0 }-2 e^{4 \epsilon_0 }-3\right)+e^{2 \epsilon_0 }\right)}{\left(e^{2 \epsilon_0 }-1\right)^2 \left(e^{2 (\sigma_0+\epsilon_0 )}+1\right)}
\Big]\,.
\end{align}

\subsection{The circular Wilson loop limit}
\label{circlimit}

We report here the $\sigma_0\to\infty$ limit of all the bosonic and fermionic determinants, representing the circular Wilson loop case $\theta_0=0$. The result for $\textrm{Det}_{\omega}\mathcal{O}_1$ in \eqref{detO1} stays obviously the same, while for the limits  of \eqref{detO2}, \eqref{detO3plus} and \eqref{detO3minus} one easily gets
\begin{flalign}\label{detO2cir}
\textrm{Det}_{\omega} \mathcal{O}_2(\theta_0=0)&=\begin{cases}
\frac{e^{\left| \omega \right|  (R-\epsilon_0)}}{2 \left| \omega \right| }& 
\,\qquad\qquad \omega\neq0\\
R&\,\qquad\qquad \omega=0
\end{cases}\\
\nonumber\\
\label{detO3pluscir}
\textrm{Det}_{\omega} \mathcal{O}_{3+}(\theta_0=0)&=\begin{cases}
\frac{e^{ (R- \epsilon_0)(\omega -1)}}{2 (\omega -1)}& 
\,\qquad\qquad \omega\geq2\\
 R&
\,\qquad\qquad \omega=1\\
 \frac{e^{ -(R- \epsilon_0)(\omega -1)}}{2(1-\omega) }&
\,\qquad\qquad \omega\leq0
\end{cases}\\
\nonumber\\
\label{detO3minuscir}
\textrm{Det}_{\omega} \mathcal{O}_{3-}(\theta_0=0)&=\begin{cases}
 \frac{e^{ (R- \epsilon_0)(\omega +1)}}{2(1 +\omega) }&
\,\qquad\qquad \omega\geq0\\
 R&
\,\qquad\qquad \omega=-1\\
-\frac{e^{ -(R- \epsilon_0)(\omega +1)}}{2 (\omega+1)}& 
\,\qquad\qquad \omega\leq-2 \,.\\
\end{cases}
\end{flalign}

The fermionic contributions \eqref{finalferm}-\eqref{specialmodes} reduce in this limit to 
\begin{align}
\textrm{Det}_{\omega}\Big[\left(\mathcal{O}_{F}^{{1},1,1}(\theta_0=0)\right)^2\Big]= \begin{cases}
\frac{e^{(R-\epsilon_0 )(2 \omega -1) }\left(\omega(e^{2 \epsilon_0 }-1)+1\right) }{4 (\omega -1) \omega ^2 \,(e^{2 \epsilon_0 }-1)}&\,\qquad\qquad\omega\geq 2 \\
\frac{R \,e^{R+\epsilon_0 }}{2 \left(e^{2 \epsilon_0 }-1\right)}&\, \qquad\qquad \omega=0,1 \\
\frac{ e^{3 (R- \epsilon_0) }\,(2 e^{2 \epsilon_0 }-1)}{16 (e^{2 \epsilon_0 }-1 )}&\, \qquad \qquad\omega=-1\,\\
\frac{e^{-(R-\epsilon_0 )(2 \omega -1) }\left((\omega -1) e^{2 \epsilon_0 }-\omega\right) }{4 (\omega -1)^2 \omega  \, (e^{2 \epsilon_0 }-1 )}&\,\qquad\qquad\omega\leq-2 \,.\\
\end{cases}
\end{align}

\section{One-loop partition functions}
\label{sec:partitionfunctions}

We now put together the determinants evaluated in the previous sections and present the one-loop partition functions for the open strings representing the latitude ($\theta_0\neq0$) and the circular ($\theta_0=0$) Wilson loop, eventually calculating their ratio.\\
In the case of fermionic determinants,  as motivated by the discussion below \eqref{fermionic_action_circle_us}, we will consider the relevant formulas \eqref{finalferm}-\eqref{specialmodes} relabelled using half-integer Fourier modes. In fact,  once projected onto the subspace labelled by $(p_{12},p_{56},p_{89})$, the spinor $\Psi$ is an eigenstate of $\Gamma_{56}$ with eigenvalue $-i p_{56}$ and the  rotation $\Psi\to\exp\left(-\frac{\tau}{2}\Gamma_{56}\right)\Psi$ reduces to a shift of the Fourier modes     by $\omega\to\omega+\frac{p_{56}}{2}$. This in particular means that below we will consider \eqref{finalferm}-\eqref{specialmodes} effectively evaluated for $\omega=s+\frac{1}{2}$ and labeled by the half-integer frequency $s$.  
In the bosonic sector -- as discussed around \eqref{bosonic_operator_product} and \eqref{detO3minus} -- we pose $\omega=\ell+1$ in $\textrm{Det}_\omega\mathcal{O}_{3+}$ together with $\omega=\ell-1$ in $\textrm{Det}_\omega\mathcal{O}_{3-}$. 
This relabeling of the frequences 
 provides in \eqref{detO3plus} and \eqref{detO3minus} a distribution of the $R$-divergences that is centered around $\ell=0$ 
({\it i.e.} with a divergence $\sim R$ for $\ell=0$ and $\sim e^{|\ell |R}$ 
for $\ell\neq 0$)  in the same way (in $\omega$)  as for the  other bosonic determinants \eqref{detO1} and \eqref{detO2}. This will  turn out to be useful while discussing  the  cancellation of $R$-dependence. 
Recalling also \eqref{classical_action}, we write the formal expression of the one-loop string action
\be\label{Zlatitude}
\!\!\!\!\!
Z({\theta_0})=e^{\sqrt{\lambda}\cos\theta_0}
\frac{\prod_{s\in\mathbb{Z}+1/2}\,\big[\,\textrm{Det}_{s}(\mathcal{O}_{F}^{{1},1,1})^2\,\textrm{Det}_{-s}(\mathcal{O}_{F}^{{1},1,1})^2\,\big]^{4/2}}
{\prod_{\ell\in\mathbb{Z}}\big[\textrm{Det}_\ell\mathcal{O}_1(\theta_0)\big]^{3/2}\,
\big[\textrm{Det}_\ell\mathcal{O}_2(\theta_0)\big]^{3/2}\,\big[\textrm{Det}_{\ell}\mathcal{O}_{3+}\left(\theta_0\right)\big]^{1/2}
\big[\textrm{Det}_{\ell}\mathcal{O}_{3-}(\theta_0)\big]^{1/2}}~\,.
\ee
To proceed, we rewrite \eqref{Zlatitude} as the (still unregularized) sum 
\begin{eqnarray}\label{sumunreg}
\Gamma({\theta_0}) &\equiv& -\log Z({\theta_0}) \equiv -\sqrt{\lambda}\cos\theta_0+\Gamma^{(1)}({\theta_0}) \\
 \Gamma^{(1)}({\theta_0}) &\equiv& \sum_{\ell\in\mathbb{Z}}\Omega_{\ell}^{B}(\theta_0)-\sum_{s\in\mathbb{Z}+1/2}\Omega_s^{F}(\theta_0)~,\nonumber
\end{eqnarray}
where the (weighted) bosonic and fermionic contributions read
\begin{eqnarray}
\Omega_{\ell}^{B}(\theta_0)&=&\frac{3}{2}\log\big[\textrm{Det}_\ell\mathcal{O}_1(\theta_0)\big]+\frac{3}{2}\log\big[\textrm{Det}_\ell\mathcal{O}_2(\theta_0)\big]+\frac{1}{2}\log\big[\textrm{Det}_{\ell}\mathcal{O}_{3+}\big]+\frac{1}{2}\log[\textrm{Det}_{\ell}\mathcal{O}_{3-}\big]\nonumber\\
\Omega_{s}^{F}(\theta_0)&=&
\frac{4}{2}\log\big[\,\textrm{Det}_{s}(\mathcal{O}_{F}^{{1},1,1})^2\big]+\frac{4}{2}\log \big[\textrm{Det}_{-s}(\mathcal{O}_{F}^{{1},1,1})^2\,\big]\,.
\end{eqnarray}
Equation \eqref{sumunreg} has the same form with effectively antiperiodic fermions encountered 
in~\cite{Kruczenski:2008zk,Dekel:2013kwa}. 
\\
Introducing the small exponential regulator $\mu$, we  proceed with the ``supersymmetric regularization'' of the one-loop effective action proposed in~\cite{Frolov:2004bh, Dekel:2013kwa}
\begin{eqnarray}\label{susyreg}
\Gamma^{(1)}(\theta_0)&=&\sum _{\ell\in \mathbb{Z}} e^{-\mu  |\ell|}\left[\Omega ^{B}_{\ell}(\theta_0)-\frac{\Omega ^{F}_{\ell+\frac{1}{2}}(\theta_0)+\Omega^{F}_{\ell-\frac{1}{2}}(\theta_0)}{2}\right]\nonumber\\
&&+\frac{\mu}{2}\Omega^{F}_{\frac{1}{2}}(\theta_0) \,+\frac{\mu}{2}  \sum _{\ell\geq 1} e^{-\mu  \ell}\left(\Omega ^{F}_{\ell+\frac{1}{2}}(\theta_0)-\Omega^{F}_{\ell-\frac{1}{2}}(\theta_0)\right)\,.
\end{eqnarray}
In the first sum (where the divergence is the same as in the original sum) one can remove  $\mu$ by sending $\mu\to0$, and use a cutoff regularization in the summation index $| \ell |\leq\Lambda$.
Importantly, the non-physical regulator $R$ disappears in \eqref{susyreg}. While in~\cite{Kruczenski:2008zk}~\footnote{In this reference  a regularization slightly different from~\cite{Frolov:2004bh, Dekel:2013kwa} was adopted.} the $R$-dependence  drops out in each summand, here it occurs as a subtle effect of the regularization scheme, and comes in the form of a cross-cancellation between the first and the second line once the sums have been carried out.
The difference in the $R$-divergence cancellation mechanism is a consequence of the different arrangement of fermionic frequencies in our regularization scheme \eqref{susyreg}.
In the circular case ($\theta_0=0$) this cancellation can be seen analytically, as in \eqref{firstsumcircle}-\eqref{othertermscircle} below. 
The same can be then inferred for the general latitude case,  since in the normalized one-loop effective action $\Gamma^{(1)}(\theta_0)-\Gamma^{(1)}(\theta_0=0)$ one observes   (see below) that the $R$-dependence drops out in each summand.

A non-trivial consistency check of \eqref{susyreg}  is to confirm that in the large $\ell$ limit the expected  UV divergences~\cite{Drukker:2000ep, Forini:2015mca}  are reproduced. Importantly, for this to happen one cannot take the limit $\epsilon_0\to0$ in the determinants above \emph{before} considering $\ell\gg1$, which is the reason why we kept  dealing with the complicated expressions for fermionic determinants above.   
Using for the Lerch transcendent in \eqref{finalferm}
\be \label{lerch_def}
\Phi(z,s,a)\equiv\sum_{n=0}^{\infty} \frac{z^n}{(n+a)^s}
\ee
the asymptotic behavior for $|a|\gg1$ ({\it i.e.} $|\ell|\gg1$ in \eqref{finalferm}) \cite{Ferreira}
\be
\Phi (z,s,a)\sim \text{sgn}(a) \left(\frac{s (s+1) z \left(z+1\right) a^{-s-2}}{2 (1-z)^3}-\frac{s\, z\, a^{-s-1}}{(1-z)^2}+\frac{a^{-s}}{1-z}\right)~,
\ee
one finds that the leading $\Lambda$-divergence is logarithmic, and - as expected from an analysis in terms of the Seeley-De
Witt coefficients~\cite{Drukker:2000ep,Forini:2015mca} -  proportional to the volume part of the Euler number 
\be\label{logdivergence}
\Gamma^{(1)}({\theta_0}) = -\chi_v(\theta_0) \sum_{1\ll|\ell |\leq\Lambda}\,\frac{1}{2 |\ell|} +O(\Lambda^0) =-\chi_v(\theta_0) \log\Lambda +O(\Lambda^0)~,\qquad\qquad\Lambda\to\infty
\ee
where
\begin{eqnarray} \label{Eulervolume}
\chi_v(\theta_0)&=&\frac{1}{4\pi} \int_{0}^{2\pi} d\tau\int_{\epsilon_0}^{\infty}d\sigma \sqrt{h} \,\, {^{(2)}\!\!\, R} \\\nonumber
&=&1-\frac{3-\cosh (2 \epsilon_0)+\cosh (2 (\sigma_0+\epsilon_0))+\cosh (2 (\sigma_0+2\epsilon_0))}{4 \sinh \epsilon_0 \cosh (\sigma_0+\epsilon_0) \cosh (\sigma_0+2\epsilon_0)}=1-\frac{1}{\epsilon}+\mathcal{O}(\epsilon)\,,
\end{eqnarray}
and we notice that this limit is independent from $\sigma_0$ ($\theta_0$).
This divergence should be cancelled via completion of the Euler number with its boundary contribution \eqref{Eulerboundary} and inclusion of the (opposite sign) measure contribution, as discussed in~\cite{Drukker:2000ep,Kruczenski:2008zk}.  
Having this in mind,  we will proceed subtracting \eqref{logdivergence} by hand in $\Gamma^{(1)}(\theta_0)$ and in $\Gamma^{(1)}(\theta_0\!=0)$.

\subsection{The circular Wilson loop}

The UV-regulated partition function in the circular Wilson loop limit reads
\begin{eqnarray}\label{susyregcircle}
\Gamma^{(1)}_{\textrm{UV-reg}}(\theta_0\!=0)&=&
\sum _{|\ell | \leq \Lambda} \left[\Omega ^{B}_{\ell}(0)-\frac{\Omega ^{F}_{\ell+\frac{1}{2}}(0)+\Omega^{F}_{\ell-\frac{1}{2}}(0)}{2} \right]+\chi_v(0) \log\Lambda\nonumber\\
&&+\frac{\mu}{2}\Omega^{F}_{\frac{1}{2}}(0) \,+\frac{\mu}{2}  \sum _{\ell\geq 1} e^{-\mu  \ell}\left(\Omega ^{F}_{\ell+\frac{1}{2}}(0)-\Omega^{F}_{\ell-\frac{1}{2}}(0)\right)\,.
\end{eqnarray}
The first line is now convergent and its total contribution evaluates for $\Lambda\to\infty$ to 
\begin{eqnarray}\label{firstsumcircle}
&&\sum_{|\ell | \leq \Lambda} \left[\Omega ^{B}_{\ell}(0)-\frac{\Omega ^{F}_{\ell+\frac{1}{2}}(0)+\Omega^{F}_{\ell-\frac{1}{2}}(0)}{2}\right]+\chi_v(0) \log\Lambda\nonumber\\
&=&\sum_{\ell=3}^{\Lambda} \log\frac{16(\ell-1)^2(\ell+1)\left(\ell+\frac{1}{\epsilon}\right)^{3}}{\ell^2 \left(2\ell+1+\frac{1}{\epsilon}\right)^2 \left(2\ell-1+\frac{1}{\epsilon}\right)^2}+\log\frac{1536 e^{-2R}\epsilon^{5/2}(1+2\epsilon)^3}{(1+3\epsilon)^4(1+5\epsilon)^2(1-\epsilon)}
+\chi_v(0) \log\Lambda\nonumber\\
&=& -2R +\log\frac{16 \,\, \Gamma\left(\frac{3}{2}+\frac{1}{2\epsilon}\right)^4}{(1-\epsilon) \,\sqrt{\epsilon} \, \Gamma\left(2+\frac{1}{\epsilon}\right)^3}~,
\end{eqnarray}
 where $\Gamma$ is Euler gamma function. The $R$-dependence in \eqref{firstsumcircle} cancels against the $\mathcal{O}(\mu^0)$ contribution stemming from the regularization-induced sum in the second line of \eqref{susyregcircle}
\begin{eqnarray}\label{othertermscircle}
\!\!\!\!\!\!
&& \frac{\mu}{2}  \sum _{\ell\geq 1} e^{-\mu  \ell}\left(\Omega ^{F}_{\ell+\frac{1}{2}}(0)-\Omega^{F}_{\ell-\frac{1}{2}}(0)\right)\nonumber \\
   &&=\mu \sum_{\ell\geq3} e^{-\ell\,\mu} \, \Big[\,2 R+\log  \frac{(\ell-1) \ell (1-\epsilon) (2\ell+1+\frac{1}{\epsilon})}
   {(\ell+1)^2 (1+\epsilon)(2\ell-1+\frac{1}{\epsilon})}\Big] \\
&&= 2R-2\,\,\textrm{arctanh}\,\epsilon\,.\nonumber
\end{eqnarray}
Summing all contributions and finally taking $\epsilon\to0$, the result is precisely as in~\cite{Kruczenski:2008zk}
\be\label{Gammacircular}
\Gamma^{(1)}_{\textrm{UV-reg}}(\theta_0\!=0)=\frac{1}{\epsilon }\left(\log \frac{\epsilon }{4} +1\right) +\frac{1}{2} \log (2 \pi )~,
\ee
despite the different frequency arrangement we commented on. We have checked that the same result is obtained employing $\zeta$-function regularization in the sum over $\ell$.  The same finite part was found in \cite{Buchbinder:2014nia}  via heat kernel methods.
There is no theoretical motivation for the  $\log\epsilon/\epsilon$-divergences appearing in  \eqref{Gammacircular}, which will be cancelled in the ratio \eqref{mainratiolog}. In~\cite{Kruczenski:2008zk}, this kind of subtraction has been done by considering the ratio between  the circular and the straight line Wilson loop.

\subsection{Ratio between latitude and circular Wilson loops}
\label{ratio}

In this section we describe the evaluation of the ratio \eqref{mainratiolog} 
\begin{eqnarray}\label{mainratioresult}
\log\frac{Z\left(\lambda,\theta_0\right)}{Z\left(\lambda,0\right)}=
\sqrt{\lambda} (\cos\theta_0-1)+
\Gamma^{(1)}_{\textrm{UV-reg}}(\theta_0\!=0)-\Gamma^{(1)}_{\textrm{UV-reg}}(\theta_0)
\end{eqnarray}
where $\Gamma^{(1)}_{\textrm{UV-reg}}(\theta_0\!=0)$ is in \eqref{susyregcircle} and $\Gamma^{(1)}_{\textrm{UV-reg}}(\theta_0)$ is regularized analogously. The complicated fermionic determinants \eqref{finalferm}-\eqref{finalfermomeganotspecial} 
make an analytical treatment highly non-trivial, and we proceed numerically.

First, we spell out \eqref{mainratioresult} as 
\begin{eqnarray}\label{gammanumerics}
\!\!\!\!\!
\Gamma^{(1)}_{\textrm{UV-reg}}(0)-\Gamma^{(1)}_{\textrm{UV-reg}}(\theta_0)
&=&
\sum _{\ell =-2}^{2}
\textstyle\left[\Omega ^{B}_{\ell}(0)-\Omega ^{B}_{\ell}(\theta_0)
-\frac{\Omega ^{F}_{\ell+\frac{1}{2}}(0)+\Omega^{F}_{\ell-\frac{1}{2}}(0)}{2} 
+\frac{\Omega ^{F}_{\ell+\frac{1}{2}}(\theta_0)+\Omega^{F}_{\ell-\frac{1}{2}}(\theta_0)}{2}\right]\nonumber\\
&+&\sum_{\ell=3}^{\Lambda}
2\!\textstyle\left[\Omega ^{B}_{\ell}(0)-\Omega ^{B}_{\ell}(\theta_0)
-\frac{\Omega ^{F}_{\ell+\frac{1}{2}}(0)+\Omega^{F}_{\ell-\frac{1}{2}}(0)}{2} 
+\frac{\Omega ^{F}_{\ell+\frac{1}{2}}(\theta_0)+\Omega^{F}_{\ell-\frac{1}{2}}(\theta_0)}{2}\right]\nonumber\\
&-&\left(\chi_v(\theta_0)-\chi_v(0)\right) \log\Lambda
+\frac{\mu}{2}\,\big[\,\Omega^{F}_{\frac{1}{2}}(0)-\Omega^{F}_{\frac{1}{2}}(\theta_0) \,\big]\\
&+&\frac{\mu}{2}  \sum _{\ell\geq 1} e^{-\mu  \ell}
\left[\Omega ^{F}_{\ell+\frac{1}{2}}(0)-\Omega^{F}_{\ell-\frac{1}{2}}(0)
-\Omega ^{F}_{\ell+\frac{1}{2}}(\theta_0)+\Omega^{F}_{\ell-\frac{1}{2}}(\theta_0)\right]\nonumber
\end{eqnarray}
where we separated the lower modes $|\ell |\leq 2$  from the sum in the second line~\footnote{This is convenient because of the different form for the special modes \eqref{specialmode1}-\eqref{specialmodes} together with the relabeling discussed above.}, and in the latter we have  used parity $\ell\to-\ell$. The sum multiplied by the small cutoff $\mu$ is zero in the limit $\mu\to0$~\footnote{This can be proved analytically since the summand behaves as $\mu \, e^{-\mu \ell} \ell^{-2}$ for large $\ell$. Removing the cutoff makes the sum vanish.}. The sum with large cutoff $\Lambda$ can be then numerically evaluated using the Euler-Maclaurin formula
\begin{eqnarray}
\sum_{\ell=m+1}^{n}f\left(\ell\right) &=&
\int_{m}^{n}f\left(\ell\right)d\ell+\frac{f\left(n\right)-f\left(m\right)}{2}+\sum_{k=1}^{p}\frac{B_{2k}}{\left(2k\right)!}\left[f^{(2k-1)}\left(n\right)-f^{(2k-1)}\left(m\right)\right]\nonumber\\
&&-\int_{m}^{n}f^{(2p)}\left(\ell\right) \: \frac{B_{2p}\left(\{\ell\} \right)}{\left(2p\right)!}d\ell \,, \qquad \qquad p\geq1\,,
\end{eqnarray}
in which $B_n(x)$ is the $n$-th Bernoulli polynomial, $B_n=B_n(0)$ is the $n$-th Bernoulli number, $\{\ell\}$ is the integer part of $\ell$, $f(\ell)$ is the summand in the second line of \eqref{gammanumerics}, so $m=2$, $n=\Lambda$. After some manipulations to improve the rate of convergence of the integrals, we safely send $\Lambda\to\infty$ in order to evaluate the normalized effective action
\begin{eqnarray} \label{gammanumerics2}
\Delta\Gamma(\theta_0)_{\textrm{sm}}&\equiv&\left[\Gamma^{(1)}_{\textrm{UV-reg}}(0)-\Gamma^{(1)}_{\textrm{UV-reg}}(\theta_0)\right]_{\textrm{sm}} 
\\\nonumber
&=&
\sum _{\ell =-2}^{2}
\left[\Omega ^{B}_{\ell}(0)-\Omega ^{B}_{\ell}(\theta_0)
-\frac{\Omega ^{F}_{\ell+\frac{1}{2}}(0)+\Omega^{F}_{\ell-\frac{1}{2}}(0)}{2} 
+\frac{\Omega ^{F}_{\ell+\frac{1}{2}}(\theta_0)+\Omega^{F}_{\ell-\frac{1}{2}}(\theta_0)}{2}\right]\nonumber\\\nonumber
&& 
 +\int_{2}^{\infty}\left[f\left(\ell\right)-\frac{\chi_{v}(\theta_0)-\chi_{v}(0)}{\ell}\right]d\ell-\left(\chi_{v}(\theta_0)-\chi_{v}(0)\right)\log2\\
&&
 -\frac{f\left(2\right)}{2}
-\sum_{k=1}^{3}\frac{B_{2k}}{\left(2k\right)!} f^{(2k-1)}\left(2\right)-\frac{1}{6!}\int_{2}^{\infty}f^{(6)}\left(\ell\right) \: B_{6}\left(\{\ell\}\right)d\ell
\nonumber\,.
\end{eqnarray}
In order to gain numerical stability for large $\ell$, above we have set $p=3$, we have cast the Lerch transcendents inside $\Omega^F_s\left(\theta_0\right)$ -- see \eqref{finalferm} -- into hypergeometric functions 
\begin{gather}
\Phi(z,1,a) = \frac{{}_2 F_1(1,a;a+1;z)}{a} \,,
\qquad\qquad |z|<1 \wedge z\neq0\, ,
\end{gather}
and we have approximated the derivatives $f^{(k)}(\ell)$ by finite-difference expressions
\begin{gather}
f^{(k)}(\ell) \,\, \to \,\, {\Delta\ell}^{-k} \sum_{i=0}^{k} (-1)^i \binom{k}{i} f \left(\ell+ (\textstyle\frac{k}{2}-i )\Delta\ell \right) \,,
\qquad \qquad \Delta\ell \ll 1~.
\end{gather}
At this stage, the expression \eqref{gammanumerics2} is only a function of the latitude parameter $\sigma_0$ ({\it i.e.} the polar angle $\theta_0$ in \eqref{theta_0_and_sigma_0}) and of two  parameters -- the IR cutoff $\epsilon_0$ and the derivative discretization $\Delta\ell$, both small compared to a given $\sigma_0$. We have tuned them in order to confidently extract four decimal digits.
\begin{figure}
    \centering
    \begin{subfigure}[b]{0.39\textwidth}
        \includegraphics[scale=0.5]{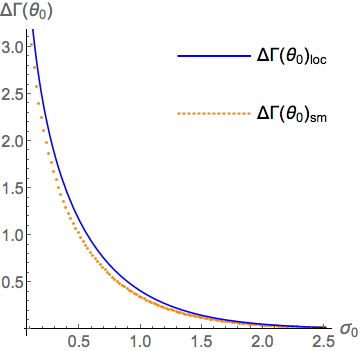}
        \caption{Comparison between $\Delta\Gamma(\theta_0)_{\rm sm}$ in \eqref{gammanumerics2} (orange dots) and $\Delta\Gamma(\theta_0)_{\rm loc}$ in \eqref{gammalocalization} (blue line).
We set $\epsilon_0=10^{-7}$, $\Delta\ell=10^{-9}$.
}
        \label{fig:Deltagamma}
    \end{subfigure}
    ~ \qquad\qquad\qquad
    \begin{subfigure}[b]{0.43\textwidth}
        \includegraphics[scale=0.5]{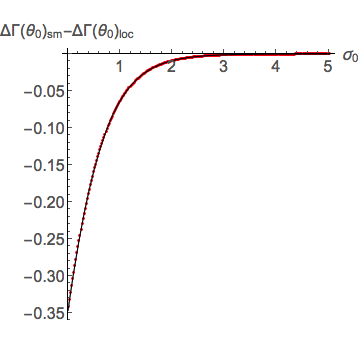}
        \caption{Fitting of the discrepancy \eqref{discrepancy} (red dots) with the test function $-\frac{1}{2}\log(1+e^{-2\sigma_0})$ (black line). We set 
        $\epsilon_0=10^{-7}$, $\Delta\ell=10^{-9}$. The interval covers approximately $ 0.8\degree \leq \theta_0 \leq 89.4\degree$.}
        \label{fig:Remainder}
    \end{subfigure}
    ~ 
     \caption{Comparison between string sigma-model perturbation theory and the predictions coming from supersymmetric localization for the ratio between latitude and circular Wilson loops in terms of the corresponding one-loop sigma-model (differences of) effective actions. }\label{fig:numerics}
\end{figure}
In Figure \ref{fig:Deltagamma} we compare the regularized one-loop effective action obtained from the perturbation theory of the string sigma-model \eqref{gammanumerics2} to the gauge theory prediction from \eqref{mainratio}
\begin{gather} \label{gammalocalization}
\Delta\Gamma(\theta_0)_{\textrm{loc}}\equiv\left[\Gamma^{(1)}_{\textrm{UV-reg}}(0)-\Gamma^{(1)}_{\textrm{UV-reg}}(\theta_0)\right]_{\textrm{loc}}= -\frac{3}{2}\log\tanh\sigma_0
\end{gather}
for different values of $\sigma_0$. Data points cover almost entirely~\footnote{When pushed to higher accuracy, numerics is computationally expensive in the vicinity of the  two limiting cases ($\sigma_0=0,\,\theta_0=\frac{\pi}{2}$) and($\sigma_0=\infty,\,\theta_0=0$).} the finite-angle region between the Zarembo Wilson loop ($\sigma_0=0,\,\theta_0=\frac{\pi}{2}$) and the circular Wilson loop ($\sigma_0=\infty,\,\theta_0=0$).

The vanishing of the normalized effective action in the large-$\sigma_0$ region is a trivial check of the normalization. As soon as  the opposite limit $\sigma_0=0$ is approached, the difference \eqref{gammanumerics2} bends up ``following'' the localization curve \eqref{gammalocalization} but also significantly deviates from it, and the measured discrepancy is incompatible with our error estimation. 
Numerics is however accurate enough to quantify the gap between the two plots on a wide range. Figure \ref{fig:Remainder} shows that, surprisingly, such gap perfectly overlaps a very simple function of $\sigma_0$ within the sought accuracy
\begin{equation}\label{discrepancy}
{\rm Rem }(\theta_0)\equiv\Delta\Gamma(\theta_0)_{\rm sm}-\Delta\Gamma(\theta_0)_{\rm loc} \approx
-\frac{1}{2} \log(1+e^{-2\sigma_0})=\log\cos\textstyle{\frac{\theta_0}{2}}~.
\end{equation}
We notice at this point that the \emph{same} simple result above can be obtained  taking in \eqref{gammanumerics} the limit of $\epsilon\to0$ \emph{before} 
performing the sums. As one can check, in this limit  UV and IR divergences cancel in the ratio~\footnote{This is also due to the volume part of the Euler number $\chi_v(\theta_0)$ being independent of $\sigma_0$ up to $\epsilon$ corrections, see \eqref{Eulervolume}.},  the  special functions in the fermionic determinants disappear and, because in general summands drastically simplify, one can proceed analytically getting the same result  calculated in terms of numerics. We remark however that such inversion of the order of sum and limit on the IR cutoff cannot be a priori justified,  as it would improperly relate the $\Lambda$ cutoff with a $1/\epsilon$ cutoff ({\it e.g.} forcing $\ell$ to be smaller than $1/\epsilon$). As emphasized above, in this limit the effective actions for the latitude and circular case separately do not reproduce  the expected UV divergences.  Therefore, the fact that in this limit the summands in the difference \eqref{gammanumerics} show a special property of convergence - which we have not analyzed in details - and lead to the exact result is a priori highly not obvious,  rendering the numerical analysis carried out in this section a rather necessary step.

On a related note, the simplicity of the result \eqref{discrepancy} and the possibility of getting an analytical result for the maximal circle $\theta_0=0$   suggest  that the summation \eqref{susyreg} could have been performed analytically also in the latitude case $\theta_0\neq0$. We have not further investigated this direction.

\section{Conclusions}
\label{sec:conclusions}

In this paper we calculated the ratio between the $\ads_5\times \sphere^5$ superstring one-loop partition functions of two supersymmetric Wilson loops  with the same topology. In so doing, we address the question whether such procedure -- which should eliminate possible ambiguities related to the measure of the partition function, under the assumption that the latter only depends on worldsheet topology -- leads to a long-sought agreement with the exact result known via localization at this order,  formula  \eqref{gammalocalization}.

Our answer is that, in the standard setup we have considered for the evaluation of the one-loop determinants (Gelfand-Yaglom approach with Dirichlet boundary conditions at the boundaries, of which one fictitious \footnote{See also Appendix~\ref{bclower} where a minimally different choice for the boundary conditions on the bosonic and fermionic modes with small Fourier mode is considered, and shown not to affect the final result. }), the agreement is not found.
%
A simple numerical fit allows us to quantify exactly a ``remainder function'', formula \eqref{discrepancy}~\footnote{See also discussion below \eqref{discrepancy}, where we notice that the same result is obtained analytically via the 
{\it a priori} not justified ``order-of-limits'' inversion.}.

 As already emphasized, the expectation that considering the ratio of string partition functions dual to Wilson loops with the same topology  should cancel measure-related ambiguities is founded on the  assumption that the partition function measure is  actually \emph{not} depending on the particular classical solution considered. Although motivated in light of literature examples similar in spirit (see Introduction), this remains an assumption, and it is not possible to exclude a priori a geometric interpretation for the observed discrepancy. 
One reasonable expectation is that the disagreement should be cured by a change of the world-sheet computational setup, tailored so to naturally lend  itself to a regularization scheme equivalent to the one (implicitly) assumed by the localization calculation~\footnote{Morally, this resembles the quest for an ``integrability-preserving'' regularization scheme, different from the most natural one suggested by worldsheet field theory considerations, in the worldsheet calculations of light-like cusps in $\mathcal{N}=4$ SYM~\cite{Giombi:2009gd} and ABJM theory~\cite{McLoughlin:2008he}.}.
%
%
%
One possibility is a choice of boundary conditions for the fermionic spectral problem~\footnote{For the bosonic sector, we do not find a reasonable alternative to the Dirichlet boundary conditions.}  different from the standard ones here adopted for the squared fermionic operator~
\footnote{For example, instead of squaring one could consider the Dirac-like first-order operator~(3.29). Then, Dirichlet boundary conditions would lead to an overdetermined system for the arbitrary integration constants of the 2 $\times$ 2 matrix-valued, first-order eigenvalue problem. The question of the non obvious alternative to consider is likely to be tied to a search of SUSY-preserving boundary conditions on the lines of~\cite{Sakai:1984vm}.}. 
 Also,  ideally  one should evaluate determinants in a diffeomorphism-preserving regularization scheme. In that it treats asymmetrically the worldsheet coordinates, the by now standard procedure of employing  the Gel'fand -Yaglom  technique for the effective (after  Fourier-transforming in $\tau$) one-dimensional case at hand  does not fall by definition in this class. In other words, the choice of using a $\zeta$-function- like regularization -- the Gel'fand-Yaglom method --  in $\sigma$ and a cutoff regularization in Fourier $\omega$-modes is a priori arbitrary.  To bypass these issues it would be desirable to fully develop a higher-dimensional formalism on the lines of~\cite{Dunne:2006ct, Kirsten:2010eg}.  A likewise fully two-dimensional method to deal with the spectral problems is the heat kernel approach, which has been employed at least for the circular Wilson loop case (where the relevant string worldsheet is the homogenous subspace $\ads_2$) in\cite{Buchbinder:2014nia,Bergamin:2015vxa}. As there explained, the procedure bypasses the need of a large $\sigma$ regulator and makes $\epsilon$ appear only in the $\ads_2$ regularized volume,  the latter being a constant multiplying the traced heat kernel and thus appearing as an overall factor in the effective action. This is different from what happens with the Gel'fand-Yaglom method, where different modes carry a different $\epsilon$-structure  and one has to identify and subtract by hand the $\epsilon$-divergence in the one-loop effective action. However, little is known about heat kernel explicit expressions for the spectra of Laplace and Dirac operators in arbitrary  two-dimensional manifolds, as it is the case as soon as the parameter $\theta_0$ is turned on. The application of the heat kernel method for the latitude Wilson loop seems then feasible only in a perturbative approach, {\it i.e.} in the small $\theta_0$ regime when the worldsheet geometry is nearly $\ads_2$~\footnote{We are grateful to A. Tseytlin for a discussion on these points.}.
It is highly desirable to address these or further possibilities in future investigations.

\section*{Acknowledgements}

We acknowledge useful  discussions with Xinyi Chen-Lin, Amit Dekel, Sergey Frolov, Simone Giombi, Jaume Gomis, Thomas Klose, Shota Komatsu, Martin Kruczenski, Daniel Medina Rincon, Diego Trancanelli, Pedro Vieira, Leo Pando Zayas, and in particular with Nadav Drukker, Arkady  Tseytlin, and  Konstantin Zarembo. We also thank A. Tseytlin and the Referee of the published version for useful comments on the manuscript.  The work of VF and EV  is funded by DFG via the Emmy Noether Programme \emph{``Gauge Field from Strings''}.  VF thanks the kind hospitality, during completion of this work, of the Yukawa Institute for Theoretical Physics in Kyoto, the Centro de Ciencias de Benasque ``Pedro Pascual", the Institute of Physics in Yerevan and in Tbilisi.  
The research of VGMP was supported in part by the University of Iceland Research Fund. 
EV acknowledges support from the Research Training Group GK 1504 \emph{``Mass, Spectrum, Symmetry''} and from the Seventh Framework Programme [FP7-People-2010-IRSES] under grant agreement n. 269217 (UNIFY), and would like to thank the Perimeter Institute for Theoretical Physics and NORDITA for hospitality during the completion of this work.
All authors would like to thank the Galileo Galilei Institute for Theoretical Physics for hospitality during the completion of this work.
\\

\appendix

\section{Notation and conventions}
\label{app:notation}

We adopt the following conventions on indices, when not otherwise stated,
\begin{gather}
\begin{tabular}{ll}
$M,N,...=0, ..., 9$ & curved target-space indices\tabularnewline
$A,B,...=0, ..., 9$ & flat target-space indices\tabularnewline
$i,j,..=0,1$ & curved worldsheet indices\tabularnewline
$a,b,..=0,1$ & flat worldsheet indices\tabularnewline
\end{tabular}
\end{gather}
Flat and curved $32\times 32$ Dirac matrices are respectively denoted by $\Gamma_A$ and $\Gamma_M$ and satisfy the $\mathfrak{so}\left(1,9\right)$
algebra
\begin{gather}
  \label{gamma_algebra_10D}
  \{\Gamma_A,\Gamma_B\}=2\eta_{AB} \mathbb{I}_{32} \qquad \{\Gamma_M,\Gamma_N\}=2 G_{MN} 
  \mathbb{I}_{32},
\end{gather} 
where $\eta_{AB}=\textrm{diag}\left(-1, +1, ..., +1\right)$ and $G_{MN}$ is the target-space metric 
(\ref{metric_new}). \\
\\
We use the explicit representation for the 10D gamma matrices
\begin{gather}
  \label{representation_gamma}
\begin{tabular}{ll}
$\Gamma_{0}=i\left(\sigma_{3}\otimes\sigma_{2}\right)\otimes\mathbb{I}_{4}\otimes\sigma_{1}$
&
$\Gamma_{5}=\mathbb{I}_{4}\otimes\left(\sigma_{3}\otimes\sigma_{2}\right)\otimes\sigma_{2}$
\tabularnewline
$\Gamma_{1}=\left(\mathbb{I}_{2}\otimes\sigma_{1}\right)\otimes\mathbb{I}_{4}\otimes\sigma_{1}$
&
$\Gamma_{6}=\mathbb{I}_{4}\otimes\left(\sigma_{1}\otimes\sigma_{2}\right)\otimes\sigma_{2}$
\tabularnewline
$\Gamma_{2}=\left(\mathbb{I}_{2}\otimes\sigma_{3}\right)\otimes\mathbb{I}_{4}\otimes\sigma_{1}$ 
&
$\Gamma_{7}=\mathbb{I}_{4}\otimes\left(-\sigma_{2}\otimes\sigma_{2}\right)\otimes\sigma_{2}$
\tabularnewline
$\Gamma_{3}=\left(\sigma_{1}\otimes\sigma_{2}\right)\otimes\mathbb{I}_{4}\otimes\sigma_{1}$
&
$\Gamma_{8}=\mathbb{I}_{4}\otimes\left(\mathbb{I}_{2}\otimes\sigma_{1}\right)\otimes\sigma_{2}$
\tabularnewline
$\Gamma_{4}=\left(-\sigma_{2}\otimes\sigma_{2}\right)\otimes\mathbb{I}_{4}\otimes\sigma_{1}$
&
$\Gamma_{9}=\mathbb{I}_{4}\otimes\left(\mathbb{I}_{2}\otimes\sigma_{3}\right)\otimes\sigma_{2}$
\tabularnewline
\end{tabular}
\end{gather}
accompanied by the chirality matrix
\begin{gather}
 \Gamma_{11}=\Gamma_{0123456789}=-\mathbb{I}_{4}\otimes\mathbb{I}_{4}\otimes\sigma_{3}.
\end{gather}
The symbol $\mathbb{I}_{n}$ stands for the $n\times n$ identity matrix and $\sigma_1,\sigma_2,\sigma_3$ 
for the Pauli matrices. It is also useful to report the combination that appears in the expansion of the fermionic 
Lagrangian (\ref{fermionic_operator_decomposed})
\begin{gather}
\Gamma_{034}=\left(\mathbb{I}_{2}\otimes\sigma_{2}\right)\otimes\mathbb{I}_{4}\otimes\sigma_{1}.
\end{gather}
The two 10D spinors of type IIB string theory have the same chirality
\begin{gather}
\label{Weyl}
\Gamma_{11} \Psi^I=\Psi^I\,, \qquad I,J=1, 2\,. 
\end{gather}
In Lorentzian signature they are subject to the Majorana condition, but this cannot be consistently imposed after Wick-rotation of the $\ads$ global time $t$. This constraint, which would halve the number of fermionic degrees of freedom, reappears as a factor $1/2$ in the exponent of fermionic determinants \eqref{Zlatitude}.

Throughout the paper we make a notational distinction between the algebraic determinant $\textrm{det}$ and the functional determinant $\textrm{Det}$, involving the determinant on the matrix indices as well as on the space spanned by ($\tau,\sigma$). We also introduce the functional determinant $\textrm{Det}_{\omega}$ over $\sigma$ for a given Fourier mode $\omega$, understanding that for any operator $\mathcal{O}$ the relation \eqref{detomega}  holds
\begin{gather}
\textrm{Det}\mathcal{O}=\prod_{\omega}\textrm{Det}_{\omega}\mathcal{O}.
\end{gather}
The boundary condition along the compact $\tau$-direction specifies if the product is over integers or half-integers. The issue related to the regularization of the infinite product is addressed in the main text. The frequencies $\omega$ label the integer modes in the Fourier-transformed bosonic and fermionic operators. We change notation and use $\ell$ for the integer and $s$ for the half-integer frequencies of the (bosonic and fermionic resp.) determinants entering the cutoff-regularized infinite products (more details in Section \ref{sec:partitionfunctions}).

Finally, a comment on the functions these matrix operators act on. They are column vectors of functions generically denoted by $\bar{f}\equiv\left(f_1, f_2, ..., f_r\right)^T$. Computing functional determinants with the techniques presented in Appendix \ref{app:gelfand_yaglom} involves solving linear differential equations, whose independent solutions $\bar{f}_{(i)}\equiv\left(f_{(i)1}, f_{(i)2}, ..., f_{(i)r}\right)^T$ are labelled by Roman numerals $i=I,II,...$.


\section{Methods for functional determinants}
\label{app:gelfand_yaglom}
The evaluation of the one-loop partition function requires the knowledge of several functional determinants of one-dimensional differential operators -- the operators in Fourier space at fixed frequency $\textrm{Det}_\omega$ (see Appendix \ref{app:notation}). This task can be  simplified via the procedure of Gel'fand and Yaglom \cite{Gelfand:1959nq} (for a pedagogical review on the topic, see \cite{Dunne:2007rt}). This algorithm has the advantage of computing ratios of determinants bypassing the computation of the full set of eigenvalues and is based on the solution of an auxiliary initial value problem~\footnote{This algorithms has been used for several examples of one-loop computations which perfectly reproduce non-trivial predictions from  ``reciprocity constraints''~\cite{Beccaria:2008tg} (see also~\cite{Beccaria:2010ry} and~\cite{Forini:2012bb}), and the general equivalence between Polyakov 
and Nambu-Goto 1-loop partition function around non-trivial solutions~\cite{Forini:2014kza}.
Further one-loop computations reproducing predictions from  quantum integrability are in~\cite{Bianchi:2013nra,Forini:2014nja,Engelund:2013fja,Roiban:2014cia}.}.

To illustrate how to proceed, let us consider the situation we typically encounter
\begin{gather}
\label{ratio_GY}
\frac{\textrm{Det}_\omega \mathcal{O}}{\textrm{Det}_\omega \hat{\mathcal{O}}}\,,
\end{gather}
in which the linear differential operators $\mathcal{O},\hat{\mathcal{O}}$ are either of first order (for fermionic degrees of freedom)~\footnote{See next section for a comment on the coincidence of the coefficient $P_0(\sigma)$ of the higher-derivative term.}
\begin{gather}
\label{prototype_first_diff_op}
\mathcal{O}=P_0(\sigma) \frac{d}{d\sigma}+P_1(\sigma)\,,
\qquad
\hat{\mathcal{O}}=P_0(\sigma) \frac{d}{d\sigma}+\hat{P}_1(\sigma)\,,
\end{gather}
or of second order (in the case of bosonic excitations)
\begin{gather}
\label{prototype_second_diff_op}
\mathcal{O}=P_0(\sigma) \frac{d^2}{d\sigma^2}+P_1(\sigma)\frac{d}{d\sigma}+P_2(\sigma)\,,
\qquad
\hat{\mathcal{O}}=P_0(\sigma) \frac{d^2}{d\sigma^2}+\hat{P}_1(\sigma)\frac{d}{d\sigma}+\hat{P}_2(\sigma).
\end{gather}
The coefficients above are complex matrices, continuous functions of $\sigma$ on the finite interval $I=\left[a,b\right]$.

In Appendix \ref{app:gelfand_yaglom_without_zero} we deal with a class of spectral problems not plagued by zero modes (vanishing eigenvalues) for chosen boundary conditions on the function space~\footnote{We mention that, for the plethora of physical situations where it is interesting to project zero modes out from the spectrum, the reader is referred to the results of \cite{Kirsten:2004qv} for self-adjoint operators of the Sturm-Liouville type as well as \cite{McKane:1995vp, Kirsten:2003py, Rajaraman} and references therein.}.
We closely follow the technology developed by Forman \cite{Forman1987, Forman1992}, who gave a prescription to work with even more general elliptic boundary value problems. We collected all the relevant formulas descending from his theorem for the bosonic sector in Appendix \ref{app:gelfand_yaglom_formulas}, and  for the square of the 2D fermionic operators in Appendix \ref{sec:square}.

Let us also stress again that the Gel'fand-Yaglom method and its extensions evaluate \emph{ratios} of determinants. Whenever we report the value of \emph{one} single determinant here and in the main text, the equal sign has to be understood up to a factor that drops out in the normalized determinant. The reference operator can be any operator with the same principal symbol.
The discrepancy can be in principle quantified for a vast class of operators with ``separated'' boundary conditions\cite{burghelea1991, burghelea1995, Lesch:1997yv}, {\it i.e.} where conditions at one boundary are not mixed with conditions at the other one.

\subsection{Differential operators of the $n$th-order}
\label{app:gelfand_yaglom_without_zero}

We consider the couple of $n$-order ordinary differential operators in one variable
\begin{gather}
\mathcal{O}=P_0(\sigma) \frac{d^n}{d\sigma^n}+\sum_{k=0}^{n-1} P_{n-k}(\sigma) \frac{d^k}{d\sigma^k}\,,
\qquad
\hat{\mathcal{O}}=P_0(\sigma) \frac{d^n}{d\sigma^n}+\sum_{k=0}^{n-1} \hat
{P}_{n-k}(\sigma) \frac{d^k}{d\sigma^k}
\end{gather}
with coefficients being $r\times r $ complex matrices. The main assumption is that the principal symbols of the two operators (proportional to the coefficient $P_0(\sigma)$ of the highest-order derivative) must be equal and invertible ($\textrm{det}P_0(\sigma)\neq0$) on the whole finite interval $I=\left[a,b\right]$. This ensures that the leading behaviour of the eigenvalues is comparable, thus the ratio is well-defined despite the fact each determinant is formally the product of infinitely-many eigenvalues of increasing magnitude. We do not impose further conditions on the matrix coefficients, besides the requirement of being continuous functions on $I$.\\
\\
The operators act on the space of square-integrable $r$-component functions $\bar{f}\equiv\left(f_1, f_2, ..., f_r\right)^T\in\mathcal{L}^2\left(I\right)$, where for our purposes one defines the Hilbert inner product ($*$ stands for complex conjugation) 
\begin{gather}
\langle \bar{f} | \bar{g} \rangle \equiv \int_a^b \Omega^2(\sigma) \sum_{i=1}^{r} f_i^*(\sigma) g_i(\sigma) d\sigma.
\end{gather}
The inclusion of the non-trivial measure factor, given by the volume element on the classical worldsheet $\sqrt{h}=\Omega^2(\sigma)$, guarantees that the worldsheet operators are self-adjoint when supplemented with appropriate boundary conditions 
\footnote{The rescaling of the operators by $\sqrt{h}$ operated in the main text removes the measure from this formula; see Appendix A in \cite{Drukker:2000ep}.}.
Indeed, to complete the characterisation of the set of functions, one specifies the $nr\times nr$ constant matrices $M,N$ implementing the linear boundary conditions at the extrema of $I$
\begin{gather}
\label{bc_GY}
M \left(\begin{array}{cc}
\bar{f}\left(a\right)\\
\frac{d}{d\sigma}\bar{f}\left(a\right)\\
\vdots\\
\frac{d^{n-1}}{d\sigma^{n-1}}\bar{f}\left(a\right)
\end{array}\right)+
N \left(\begin{array}{cc}
\bar{f}\left(b\right)\\
\frac{d}{d\sigma}\bar{f}\left(b\right)\\
\vdots\\
\frac{d^{n-1}}{d\sigma^{n-1}}\bar{f}\left(b\right)
\end{array}\right)=
\left(\begin{array}{cc}
0\\
0\\
\vdots\\
0
\end{array}\right).
\end{gather}
\\
The particular significance of the Gel'fand-Yaglom theorem and its extensions, specialized in~\cite{Forman1987, Forman1992}  to elliptic differential operators, lies in the fact that it astonishingly cuts down the complexity of finding the spectrum of the operators of interests
\begin{gather}
\mathcal{O} \bar{f}_{\lambda}(\sigma)=\lambda \bar{f}_{\lambda}(\sigma)\,,
\quad\qquad
\hat{\mathcal{O}} \hat{\bar{f}}_{\hat{\lambda}}(\sigma)=\hat{\lambda} \hat{\bar{f}}_{\hat{\lambda}}(\sigma),
\end{gather}
and then finding a meromorphic extension of $\zeta$-function. All this is encoded into the elegant formula
\begin{gather}
\label{ratio_forman}
\frac{\textrm{Det}_\omega \mathcal{O}}{\textrm{Det}_\omega \hat{\mathcal{O}}}=
\frac{\exp\left\{\int_a^b \textrm{tr} \left[\mathcal R(\sigma) P_1(\sigma) P_0^{-1}(\sigma)\right] d\sigma\right\} \textrm{det}\left[M+N Y_{\mathcal{O}}\left(b\right)\right]}
{\exp\left\{\int_a^b \textrm{tr} \left[\mathcal R(\sigma) \hat{P}_1(\sigma) P_0^{-1}(\sigma)\right] d\sigma\right\} \textrm{det}\left[M+N Y_{\hat{\mathcal{O}}}\left(b\right)\right]}\,,
\end{gather}
for the ratio (\ref{ratio_GY}), and where $\mathcal R$ is defined below. 
This result agrees with the  one obtained via $\zeta-$function regularization for elliptic differential operators. Notice that any constant rescaling of $M,N$ in (\ref{bc_GY}) leaves the ratio unaffected. Moreover, if also the next-to-higher-derivative coefficients coincide ($P_1(\sigma)=\hat{P}_1(\sigma)$), the exponential factors cancel out.
The $nr\times nr$ matrix
\begin{gather}
\label{GY_Y}
Y_{\mathcal{O}}(\sigma)=\left(\begin{array}{cccc}
\bar{f}_{(I)}(\sigma) & \bar{f}_{(II)}(\sigma) & \dots & \bar{f}_{(nr)}(\sigma)\\
\frac{d}{d\sigma}\bar{f}_{(I)}(\sigma) & \frac{d}{d\sigma}\bar{f}_{(II)}(\sigma) & \dots & \frac{d}{d\sigma}\bar{f}_{(nr)}(\sigma)\\
\vdots & \vdots & \ddots & \vdots \\
\frac{d^{n-1}}{d^{n-1}\sigma}\bar{f}_{(I)}(\sigma) & \frac{d^{n-1}}{d^{n-1}\sigma}\bar{f}_{(II)}(\sigma) & \dots & \frac{d^{n-1}}{d^{n-1}\sigma}\bar{f}_{(nr)}(\sigma)\\
\end{array}\right)
\end{gather}
accommodates all the independent homogeneous solutions of
\begin{gather}
\mathcal{O}\bar{f}_{(i)}(\sigma)=0 \qquad i=I, II, ..., 2r
\end{gather}
chosen such that $Y_{\mathcal{O}}\left(a\right)=\mathbb{I}_{nr}$. It can be thought of as the fundamental matrix of the equivalent first-order operator acting on $nr$-tuples of functions. $Y_{\hat{\mathcal{O}}}(\sigma)$ is similarly defined with respect to $\hat{\mathcal{O}}$.\\
\\
If we restrict  to even-order differential operators, then $\mathcal R(\sigma)=\frac{1}{2} \mathbb{I}_{nr}$ and (\ref{ratio_forman}) simplifies:
\begin{gather}
\label{ratio_forman_even}
\frac{\textrm{Det}_\omega \mathcal{O}}{\textrm{Det}_\omega \hat{\mathcal{O}}}=
\frac{\exp\left\{\frac{1}{2} \int_a^b \textrm{tr} \left[P_1(\sigma) P_0^{-1}(\sigma)\right] d\sigma\right\} \textrm{det}\left[M+N Y_{\mathcal{O}}\left(b\right)\right]}
{\exp\left\{\frac{1}{2} \int_a^b \textrm{tr} \left[\hat{P}_1(\sigma) P_0^{-1}(\sigma)\right] d\sigma\right\} \textrm{det}\left[M+N Y_{\hat{\mathcal{O}}}\left(b\right)\right]}\,.
\end{gather}
For odd $n$ one gets a slightly more complicated structure, constructed as follows. Let us assume that the (generalized) spectrum of the principal symbol of $\mathcal{O},\hat{\mathcal{O}}$, {\it i.e.} the matrix $\left(-i\right)^n P_0(\sigma)$, has no intersection with the cone $C\equiv\{z\in\mathbb{C}| \bar{\theta}_1<\textrm{arg}{z}<\bar{\theta}_2\}$  for some choice of $\bar{\theta}_1,\bar{\theta}_2$. This is to say that $\mathcal{O}$ has principal angle between $\bar{\theta}_1$ and $\bar{\theta}_2$. It also follows that no eigenvalue falls in the opposite cone $-C\equiv\{z\in\mathbb{C}| \bar{\theta}_1+\pi<\textrm{arg}{z}<\bar{\theta}_2+\pi\}\}$ when $n$ is odd. Consequently, the finitely-many eigenvalues fall under two sets, depending on which sector of $\mathbb{C}\setminus\left(C\cup-C\right)$ they belong to. The matrix $\mathcal R(\sigma)$ is then defined~\footnote{Up to a factor $\frac{1}{n}$, see amendment in \cite{Forman1992}.} as the projector onto the subspace spanned by the eigenvectors corresponding to all eigenvalues in one of these two subsets of the complex plane.\\
\\
We did not use this formula for  odd $n$  in this paper, but  notice that this machinery could be potentially applied to the first-order fermionic \eqref{O111} operator.

\subsection{Applications}
\label{app:gelfand_yaglom_formulas}
We list the applications of the theorem (\ref{ratio_forman}) for the scalar-/matrix-valued operators in the main text. 
In the following we leave out formulas for hatted operators and solutions in order not to clutter formulas, understanding that they satisfy the same initial value problems.

\begin{itemize}
\item Second-order scalar-valued differential operators    $\mathcal{O}_1$, $\mathcal{O}_2\left(\theta_0\right)$,  $\mathcal{O}_{3\pm}\left(\theta_0\right)$,\\
 Dirichlet boundary conditions $f_1\left({\epsilon_0}\right)=f_1\left(R\right)=0$.
\begin{gather}
\label{O2first}
M=
\left(\begin{array}{cc}
1 & 0\\
0 & 0
\end{array}\right)
\qquad
N=
\left(\begin{array}{cc}
0 & 0\\
1 & 0
\end{array}\right)
\qquad
\frac{\textrm{Det}_{\omega}\left[\frac{d^2}{d\sigma^2}+P_2(\sigma)\right]}
{\textrm{Det}_{\omega}\left[\frac{d^2}{d\sigma^2}+\hat{P}_2(\sigma)\right]}
=
\frac{f_{(II)1}\left(R\right)}{\hat{f}_{(II)1}\left(R\right)}
\end{gather}
The normalization of the matrix (\ref{GY_Y}) tells that the function $f_{(II)1}\left(\sigma\right)$ solves the initial value problem
\begin{gather}\label{O2second}
f_{(II)1}^{\prime\prime}(\sigma)+P_2(\sigma)f_{(II)1}(\sigma)=0
\qquad
f_{(II)1}\left({\epsilon_0}\right)=0
\qquad
f_{(II)1}^{\prime}\left({\epsilon_0}\right)=1.
\end{gather}

\item Second-order $2\times2$ matrix-valued differential operators $\mathcal{O}_3\left(\theta_0\right)$,\\
 Dirichlet boundary conditions $f_1\left({\epsilon_0}\right)=f_2\left({\epsilon_0}\right)=f_1\left(R\right)=f_2\left(R\right)=0$.
\begin{equation}
\label{O3first}
\begin{split}
&M=
\left(\begin{array}{cccc}
1 & 0 & 0 & 0\\
0 & 1 & 0 & 0\\
0 & 0 & 0 & 0\\
0 & 0 & 0 & 0
\end{array}\right)
\qquad
N=
\left(\begin{array}{cccc}
0 & 0 & 0 & 0\\
0 & 0 & 0 & 0\\
1 & 0 & 0 & 0\\
0 & 1 & 0 & 0
\end{array}\right) 
\\
&\frac{\textrm{Det}_{\omega}\left[\frac{d^2}{d\sigma^2}+P_2(\sigma)\right]}
{\textrm{Det}_{\omega}\left[\frac{d^2}{d\sigma^2}+\hat{P}_2(\sigma)\right]}
=
\frac{f_{(III)1}\left(R\right)f_{(IV)2}\left(R\right)-f_{(III)2}\left(R\right)f_{(IV)1}\left(R\right)}
{\hat{f}_{(III)1}\left(R\right)\hat{f}_{(IV)2}\left(R\right)-\hat{f}_{(III)2}\left(R\right)\hat{f}_{(IV)1}\left(R\right)}
\end{split}
\end{equation}
where
\begin{equation}
\label{O3second}
\begin{split}
&\left(\begin{array}{c}
 f_{(III)1}^{\prime\prime}(\sigma)\\
 f_{(III)2}^{\prime\prime}(\sigma)
\end{array}\right)
+P_2(\sigma)
\left(\begin{array}{c}
 f_{(III)1}(\sigma)\\
 f_{(III)2}(\sigma)
\end{array}\right)=
\left(\begin{array}{c}
0\\
0
\end{array}\right)\\
&f_{(III)1}\left({\epsilon_0}\right)=f_{(III)2}\left({\epsilon_0}\right)=f_{(III)2}^\prime\left({\epsilon_0}\right)=0
\qquad
f_{(III)1}^\prime\left({\epsilon_0}\right)=1 \\
\\
&\left(\begin{array}{c}
 f_{(IV)1}^{\prime\prime}(\sigma)\\
 f_{(IV)2}^{\prime\prime}(\sigma)
\end{array}\right)
+P_2(\sigma)
\left(\begin{array}{c}
 f_{(IV)1}(\sigma)\\
 f_{(IV)2}(\sigma)
\end{array}\right)=
\left(\begin{array}{c}
0\\
0
\end{array}\right) \\
&f_{(IV)1}\left({\epsilon_0}\right)=f_{(IV)2}\left({\epsilon_0}\right)=f_{(IV)1}^\prime\left({\epsilon_0}\right)=0
\qquad
f_{(IV)2}^\prime\left({\epsilon_0}\right)=1 ~.
\end{split}
\end{equation}

\item Second-order $2\times2$ matrix-valued differential operators $\left[\mathcal{O}^{p_{12},p_{56},p_{89}}_{F}\left(\theta_0\right)\right]^2$,\\
 Dirichlet boundary conditions $f_1\left({\epsilon_0}\right)=f_2\left({\epsilon_0}\right)=f_1\left(R\right)=f_2\left(R\right)=0$.
\begin{gather}\label{OF2first}
M=
\left(\begin{array}{cccc}
1 & 0 & 0 & 0\\
0 & 1 & 0 & 0\\
0 & 0 & 0 & 0\\
0 & 0 & 0 & 0
\end{array}\right)
\qquad
N=
\left(\begin{array}{cccc}
0 & 0 & 0 & 0\\
0 & 0 & 0 & 0\\
1 & 0 & 0 & 0\\
0 & 1 & 0 & 0
\end{array}\right)
\qquad
Y_{\mathcal{O}}\left(\sigma\right)=
\left(\begin{array}{cc}
f_{(I)1}\left(\sigma\right) & f_{(II)1}\left(\sigma\right) \\
f_{(I)2}\left(\sigma\right) & f_{(II)2}\left(\sigma\right)
\end{array}\right)
\\
\frac
{\textrm{Det}_\omega\left[P_0\left(\sigma\right)\frac{d}{d\sigma}+P_1\left(\sigma\right)\right]^2}
{\textrm{Det}_\omega\left[P_0\left(\sigma\right)\frac{d}{d\sigma}+\hat{P}_1\left(\sigma\right)\right]^2}=
\frac{\int_{\epsilon_0}^R ds Y^{-1}_{\mathcal{O}}\left(s\right) P^{-1}_0\left(s\right)Y_{\mathcal{O}}\left(s\right)}
{\int_{\epsilon_0}^R ds Y^{-1}_{\mathcal{\hat{O}}}\left(s\right) P^{-1}_0\left(s\right)Y_{\mathcal{\hat{O}}}\left(s\right)}
\end{gather}
with 
\begin{gather}
P_0(\sigma)
\left(\begin{array}{c}
 f_{(I)1}^{\prime}(\sigma)\\
 f_{(I)2}^{\prime}(\sigma)
\end{array}\right)
+P_1(\sigma)
\left(\begin{array}{c}
 f_{(I)1}(\sigma)\\
 f_{(I)2}(\sigma)
\end{array}\right)=
\left(\begin{array}{c}
0\\
0
\end{array}\right)
\qquad
f_{(I)1}\left({\epsilon_0}\right)=1
\qquad
f_{(I)2}\left({\epsilon_0}\right)=0\nonumber\\
\\
P_0(\sigma)
\left(\begin{array}{c}
 f_{(II)1}^{\prime}(\sigma)\\
 f_{(II)2}^{\prime}(\sigma)
\end{array}\right)
+P_1(\sigma)
\left(\begin{array}{c}
 f_{(II)1}(\sigma)\\
 f_{(II)2}(\sigma)
\end{array}\right)=
\left(\begin{array}{c}
0\\
0
\end{array}\right)
\qquad
f_{(II)1}\left({\epsilon_0}\right)=0
\qquad
f_{(II)2}\left({\epsilon_0}\right)=1 \,.\nonumber
\end{gather}
This is a corollary of (\ref{op_square_Dirichlet}).
\end{itemize}

\subsection{Square of first-order differential operators}
\label{sec:square}
As a consequence of the Forman's construction, we can easily compute the ratio of determinants of the square of first-order operators with reference only to the operators themselves. Consider the matrix operator of the form (\ref{prototype_first_diff_op}) 
\footnote{We omit to report similar formulas for the hatted operator $\hat{\mathcal{O}}=P_0\left(\sigma\right)\frac{d}{d\sigma} + \hat{P}_1\left(\sigma\right)$.}
\begin{gather}
\mathcal{O}=P_0\left(\sigma\right)\frac{d}{d\sigma} + P_1\left(\sigma\right)
\end{gather}
and denote by $Y_{\mathcal{O}}\left(\sigma\right)$ its fundamental matrix, which solves the equation (here $'$ is the derivative with respect to $\sigma$)
\begin{gather}
P_0\left(\sigma\right) Y^{'}_{\mathcal{O}}\left(\sigma\right) + P_1\left(\sigma\right) Y_{\mathcal{O}}\left(\sigma\right)=0, \qquad Y_{\mathcal{O}}\left(a \right)=\mathbb{I}_r.
\end{gather}
The matrix of fundamental solutions of the square of this operator
\begin{gather}
\mathcal{O}^2=P^2_0\left(\sigma\right)\frac{d^2}{d\sigma^2} + \left[P_0\left(\sigma\right)P^{'}_0\left(\sigma\right)+\left\{P_0\left(\sigma\right),P_1\left(\sigma\right) \right\}\right] \frac{d}{d\sigma}+P^2_1\left(\sigma\right)+P_0\left(\sigma\right)P^{'}_1\left(\sigma\right)
\end{gather}
can be constructed via the method of reduction of order as
\begin{gather}
\label{OY}
Y_{\mathcal{O}^2}\left(\sigma\right)=
\left(\begin{array}{cc}
Y_{\mathcal{O}}\left(\sigma\right)-Z\left(\sigma\right) Y^{'}_{\mathcal{O}}\left(a\right) & \qquad Z\left(\sigma\right)\\
Y^{'}_{\mathcal{O}}\left(\sigma\right)-Z^{'}\left(\sigma\right) Y^{'}_{\mathcal{O}}\left(a\right) & \qquad Z^{'}\left(\sigma\right)
\end{array}\right),
\qquad Y_{\mathcal{O}^2}\left(a\right)=\mathbb{I}_{2r}
\end{gather}
in which
\begin{gather}
\label{Z}
Z\left(\sigma\right)=Y_{\mathcal{O}}\left(\sigma\right) \int_a^b ds \left[Y^{-1}_{\mathcal{O}}\left(s\right) P^{-1}_0\left(s\right)Y_{\mathcal{O}}\left(s\right)\right] P_0\left(a\right) \qquad Z\left(a\right)=0 \qquad Z'\left(a\right)=\mathbb{I}_r.
\end{gather}
encapsulates the solutions of $\mathcal{O}\bar{f}=0$ and two more ones of $\mathcal{O}^2 \bar{f}=0$.\\
\\
Suppose that the spectral problem of the squared operator is determined by the boundary condition
\begin{gather} \label{DDfermbc}
M_{\mathcal{O}^2} \bar{f}\left(a\right)+N_{\mathcal{O}^2} \bar{f}\left(b\right)=0\,.
\end{gather}
After some algebra, successive applications of (\ref{ratio_forman_even}),(\ref{OY}),(\ref{Z}) bring
\begin{gather}
\label{op_square}
\textrm{Det}_\omega\mathcal{O}^2 =
\sqrt{\frac{\textrm{det}P_0\left(b\right)}{\textrm{det}P_0\left(a\right)}}
\frac{\textrm{det}\left[M_{\mathcal{O}^2}+N_{\mathcal{O}^2}Y_{\mathcal{O}^2}\left(b\right)\right]}
{\textrm{det}Y_{\mathcal{O}}\left(b\right)} \,.
\end{gather}
For Dirichlet boundary conditions at both endpoints $\sigma=a,\,b$ used in the present paper
\begin{gather} \label{DDfermMN}
f_1\left({a}\right)=f_2\left({a}\right)=f_1\left(b\right)=f_2\left(b\right)=0  \\
M_{\mathcal{O}^2}=
\left(\begin{array}{cc}
\mathbb{I}_r & 0\\
0 & 0
\end{array}\right)
\qquad
N_{\mathcal{O}^2}=
\left(\begin{array}{cc}
0 & 0\\
\mathbb{I}_r & 0
\end{array}\right) \nonumber
\end{gather}
then (\ref{op_square}) gives
\begin{gather}
\label{op_square_Dirichlet}
\left(\textrm{Det}_\omega \mathcal{O}^2\right)_\textrm{Dirichlet}=
\sqrt{\textrm{det}P_0\left(a\right) \textrm{det}P_0\left(b\right)}
~~
\textrm{det}\left[\int_a^b ds Y^{-1}_{\mathcal{O}}\left(\sigma\right) P^{-1}_0\left(\sigma\right)Y_{\mathcal{O}}\left(\sigma\right)\right].
\end{gather}

\section{Fermionic determinant $\textrm{Det}\mathcal{O}_{F}\left(\theta_0\right)$: details}
\label{app:detailsferm}

In this Appendix we collect some details on the analysis of the fermionic determinant $\textrm{Det}\mathcal{O}_{F}\left(\theta_0\right)$  in \eqref{prod_fermionic_operator_decomposed}.

\subsection{Derivation of  \eqref{detfermall1mirror}}
\label{ferm111}

To begin our analysis of $\textrm{Det}\mathcal{O}_{F}\left(\theta_0\right)$ in \eqref{prod_fermionic_operator_decomposed}, let us observe that  
\be
\label{equiv1}
\Gamma^{34}\mathcal{O}_{F}^{p_{12},p_{56},p_{89}}\Gamma^{43}=-\mathcal{O}_{F}^{-p_{12},p_{56},p_{89}}~.
\ee
This fact can be used to show that  
\be\label{property1}
\textrm{Det}(\mathcal{O}_{F}^{1,p_{56},p_{89}} )\textrm{Det}(\mathcal{O}_{F}^{-{1},p_{56},p_{89}} )=\textrm{Det}[(\mathcal{O}_{F}^{{1},p_{56},p_{89}} )^2].
\ee
 Indeed, let us denote the ``positive" eigenvalues of $\mathcal{O}_{F}^{{1},p_{56},p_{89}} $ with $\{\lambda_n , \, \mathrm{Re}(\lambda_n)>0\}$ and 
 the ``negative" ones with  $\{-\mu_n , \, \mathrm{Re}(\mu_n)>0\}$.  Because of the relation \eqref{equiv1}, the spectrum of    $\mathcal{O}_{F}^{-{1},p_{56},p_{89}} $  is given by 
 \begin{equation}
 \{-\lambda_n\}\cup \{\mu_n\} \,.
 \end{equation}
 The $\zeta-$function for the first operator  is 
 \be
 \zeta^{1,p_{56},p_{89}}(s)=\sum_{n}(\lambda_n)^{-s}+ e^{\mp i \pi s}\sum_{n}(\mu_n)^{-s},
 \ee
 while for the second operator we find 
  \be
 \zeta^{-1,p_{56},p_{89}}(s)= e^{\mp i \pi s}\sum_{n}(\lambda_n)^{-s}+\sum_{n}(\mu_n)^{-s}.
 \ee
 Summing the two contributions we obtain
 \be
 \zeta^{1,p_{56},p_{89}}(s)+ \zeta^{-1,p_{56},p_{89}}(s)=(1+ e^{\mp i \pi s})\left [\sum_{n}(\lambda_n)^{-s}+ \sum_{n}(\mu_n)^{-s}\right]\equiv(1+ e^{\mp i \pi s})\Xi (s)
 \ee 
 The spectrum of $(\mathcal{O}_{F}^{1,p_{56},p_{89}})^2$ is  given  instead by
 \begin{gather}
 \{\lambda_n^2\}\cup\{\mu_n^2\}
 \end{gather}
  and the corresponding $\zeta-$function is 
 \be\label{ZetaXi}
 Z(s)=\left [\sum_{n}(\lambda^2_n)^{-s}+ \sum_{n}(\mu^2_n)^{-s}\right]=\Xi(2 s).
 \ee
 Therefore 
\begin{eqnarray}
 \log\big[\textrm{Det}(\mathcal{O}_{F}^{1,p_{56},p_{89}} )\textrm{Det} (\mathcal{O}_{F}^{-1,p_{56},p_{89}} )\big]&=& 
 -\frac{d \zeta^{1,p_{56},p_{89}}(0)}{ds}-\frac{d\zeta^{-1,p_{56},p_{89}}(0)}{ds}\nonumber\\
&=&\pm i \pi \Xi(0)-2\Xi^\prime(0)\,,\\
\log(\textrm{Det}[(\mathcal{O}_{F}^{p_{12},p_{56},p_{89}} )^2])&=&-2\Xi^\prime(0)\,,\nonumber
\end{eqnarray}
so that it holds
\be
\label{cirone}
 \log(\textrm{Det}(\mathcal{O}_{F}^{1,p_{56},p_{89}} )\textrm{Det} (\mathcal{O}_{F}^{-{1},p_{56},p_{89}} ))=
 \log(\textrm{Det}[(\mathcal{O}_{F}^{{1},p_{56},p_{89}} )^2])\pm i \pi \Xi(0)~.
\ee
We can namely express the combination $\textrm{Det}(\mathcal{O}_{F}^{1,p_{56},p_{89}} )\textrm{Det} (\mathcal{O}_{F}^{-{1},p_{56},p_{89}} )$ 
in terms of determinant and  the $\zeta$-function in $0$ ($\Xi(0)$)  of the \emph{squared} operator 
$(\mathcal{O}_{F}^{{1},p_{56},p_{89}} )^2$. \\
We now use Corollary 2.4 of \cite{burghelea1995}
\be
Z(0)=r\left(\frac{|\alpha|+|\beta|}{2n}-n+1\right)\,,
\ee
where $2n$ is the order of the differential operator, $\alpha,\beta$ are 
parameters that only depend on the boundary conditions and $r$ is the matrix dimension of the operators. In our case 
($n=1$,\,$|\alpha|=|\beta|=1$,\,$r=2$) it is  $Z(0)=2$ and thus via \eqref{ZetaXi} $\Xi(0)=2$, 
to conclude that 
\be
\textrm{Det}(\mathcal{O}_{F}^{1,p_{56},p_{89}} )\textrm{Det} (\mathcal{O}_{F}^{-{1},p_{56},p_{89}} )
= \textrm{Det}[(\mathcal{O}_{F}^{{1},p_{56},p_{89}} 
)^2]\,,
\ee
and thus \eqref{property1} is proven.\\
\\
The determinant of the fermionic operator can then be written as~\footnote{The Corollary 2.4~\cite{burghelea1995} 
can be easily check to hold both for $\zeta_{-1,1}(0)$ and for $\zeta_{1,1}(0)$.}
\begin{eqnarray}\nonumber
\textrm{Det}[\mathcal{O}_{F}\left(\theta_0\right)]&=&
\underset{p_{12},p_{56},p_{89}=-1,1}{\prod}\textrm{Det} 
[\mathcal{O}_{F}^{p_{12},p_{56},p_{89}}\left(\theta_0\right)]\\\label{fermeven}
&=&\textrm{Det}[(\mathcal{O}_{F}^{{1},1,1} )^2]^2 \textrm{Det}[(\mathcal{O}_{F}^{{1},-1,1} )^2]^2,
\end{eqnarray}
where we have used the property \eqref{cirone} and that the operators   $\mathcal{O}_{F}^{p_{12},p_{56},p_{89}}$ in \eqref{fermionic_operator_decomposed}
do not depend  on the value of $p_{89}$. \\
\\
We can also easily argue that   $\textrm{Det}[(\mathcal{O}_{F}^{{1},-1,1} )^2]=\textrm{Det}[(\mathcal{O}_{F}^{{1},1,1} )^2]$. 
Let $\{\lambda_n, \psi(\tau,\sigma)\}$  be the spectrum of  $\mathcal{O}_{F}^{{1},1,1}$, then $\{-\lambda_n, \Gamma_4 \psi (-\tau,\sigma))\}$ is
the spectrum of   $\mathcal{O}_{F}^{{1},-1,1}$. Indeed it is
\begin{eqnarray}
\mathcal{O}_{F}^{1,-1,1} \Gamma_4  \psi(-\tau,\sigma)
&\equiv&
\frac{i}{\Omega(\sigma)}\left(\Gamma_{4}\partial_{\tau}+\Gamma_{3}\partial_{\sigma}-a_{34}(\sigma)\Gamma_{3}+i a_{56}(\sigma)\Gamma_{4}\right)
 \Gamma_4  \psi(-\tau,\sigma)\nonumber\\\label{p56even}
&&+\frac{1}{\Omega^{2}(\sigma)}\left(-i\sinh^{2}\rho(\sigma)\Gamma_{0}+\sin^{2}\theta(\sigma)\Gamma_{034}\right)\Gamma_4  
\psi(-\tau,\sigma)\nonumber\\\nonumber
&=&- \Gamma_4\left[\frac{i}{\Omega(\sigma)}\left(\Gamma_{4}\partial_{(-\tau)}+\Gamma_{3}\partial_{\sigma}-a_{34}(\sigma)\Gamma_{3}-i a_{56}(\sigma)\Gamma_{4}\right)
\psi(-\tau,\sigma)\right.\\\nonumber
&&+\left.\frac{1}{\Omega^{2}(\sigma)}\left(-i\sinh^{2}\rho(\sigma)\Gamma_{0}-\sin^{2}\theta(\sigma)\Gamma_{034}\right)  
\psi(-\tau,\sigma)\right]\\\label{mirroromega}
&\equiv&- \Gamma_4\,\mathcal{O}_{F}^{1,1,1}\,\psi(-\tau,\sigma)
=-\lambda_n \,\Gamma_4\,\psi_1(-\tau,\sigma). 
\end{eqnarray}
Thus the  eigenvalues of the \emph{squared} operator $(\mathcal{O}_{F}^{{1},-1,1} )^2$ 
are the same of those of the \emph{squared} operator $(\mathcal{O}_{F}^{{1},1,1} )^2$ 
and consequently  the two determinants  coincide.
In restricting ourselves to one-dimensional spectral problems - and thus 
working in terms of Fourier modes $\omega$ and referring to \eqref{detomega} - from the statement \eqref{mirroromega}
one obtains 
\be 
\textrm{Det}_{\omega}[(\mathcal{O}_{F}^{{{1},-1,1}})^2]= 
\textrm{Det}_{-\omega}[(\mathcal{O}_{F}^{{1},1,1} )^2]~.
\ee
from which  \eqref{detfermall1mirror} follows.
 
 \subsection{Simplifying the large-$R$ expression for $\textrm{Det}_\omega (\mathcal{O}_{F}^{{1},1,1})^2$ }
\label{largeRsimplified}
As from \eqref{op_square}, the key-ingredient in the explicit computation of $\textrm{Det}_\omega (\mathcal{O}_{F}^{{1},1,1} )^2$ \eqref{O111} is  $Y(\sigma)$, the $2\times 2$ matrix  of the fundamental solutions obeying the boundary conditions 
$ Y(\epsilon_0)=\mathbb{I}_2$, as in \eqref{OY}. It is not difficult to explicitly check that the structure  of this matrix can be parametrized as follows
\be
 Y(\sigma)= e^{\omega(\sigma-{\epsilon_0})}S_1(\sigma)+ e^{-\omega(\sigma-{\epsilon_0})}S_2(\sigma),
\ee
where  the entries of the matrices $S_1(\sigma)$ and $S_2(\sigma)$ depends on $\omega$ only  through rational functions.   We can infer some important properties of these matrices from the fact  that  $Y(\sigma)$ obeys its  secular equation
\be
\label{CaleyHamilton}
Y^2 (\sigma)-\mathrm{tr}(Y(\sigma)) Y(\sigma)+\textrm{det} (Y(\sigma)){\mathbb{I}_2}=0\,.
\ee
In particular  one can easily check that $\textrm{det} Y(\sigma)$ does not depend on $\omega$.   Then  the secular equation becomes the following equation for $S_1$ and $S_2$ \footnote{We omit the $\sigma$-dependence in the matrices.}
\begin{eqnarray}\nonumber
&&e^{2\omega(\sigma-{\epsilon_0})} (S_1^2-\mathrm{tr}(S_1) S_1)+ e^{-2\omega(\sigma-{\epsilon_0})}(S_2^2-\mathrm{tr}(S_2)S_2)\\
&&
+ \{S_1,S_2\}
-\mathrm{tr}(S_1) S_2
-\mathrm{tr}(S_2) S_1 +\textrm{det} (Y){\mathbb{I}_2}=0
\end{eqnarray}
and therefore
\be
\label{cipro}
S_1^2-\mathrm{tr}(S_1) S_1=0\,,\ \ \ \ \  S_2^2-\mathrm{tr}(S_2)S_2=0\,,\ \ \ \ \  \{S_1,S_2\}
-\mathrm{tr}(S_1) S_2
-\mathrm{tr}(S_2) S_1 +\textrm{det} (Y){\mathbb{I}_2}=0\,.
\ee
The  matrices $S_1$ and $S_2$ must also satisfy the differential equations 
\be
P_0 \partial_\sigma  S_1 +(P_1+\omega P_0) S_1=0\,,\ \ \\ \ \ \ \ 
P_0 \partial_\sigma  S_2 +(P_1-\omega P_0) S_2=0\,,
\ee
where $P_0, P_1$ appear in the Dirac-like operator $\mathcal{O}_{F}^{{1},1,1} $ written in the form $P_0 \partial_\sigma +P_1$. 
Since $P_0$ is invertible we can symbolically write  this  as ${S'_i}+M_i S_i=0$.  This implies a set of interesting properties:
\begin{eqnarray}
(S_i)_{1,2}({S'}_i)_{1,1} -(S_i)_{1,1}({S'}_i)_{1,2} &=&-  (S_i)_{1,2}[ (M_i)_{1,1} (S_i)_{1,1}+(M_i)_{1,2} (S_i)_{2,1}]\\\nn
&+&(S_i)_{1,1}[ (M_i)_{1,1} (S_i)_{1,2}+(M_i)_{1,2} (S_i)_{2,2}]=(M_i)_{1,2} \textrm{det}(S_i)=0.
\end{eqnarray}
and
\begin{eqnarray}
(S_i)_{21}({S'}_i)_{2,2} -(S_i)_{2,2}({S'}_i)_{2,1} &=&-  (S_i)_{2,1}[ (M_i)_{2,1} (S_i)_{1,2}+(M_i)_{2,2} (S_i)_{2,2}]\\\nn
&+&(S_i)_{2,2}[ (M_i)_{2,1} (S_i)_{1,1}+(M_i)_{2,2} (S_i)_{2,1}]=(M_i)_{2,1} \textrm{det}(S_i)=0.
\end{eqnarray}
Namely the ratios
\be
\frac{(S_i)_{11}}{(S_i)_{12}} \ \ \ \ \ \mathrm{and}\ \ \ \ \  \frac{(S_i)_{21}}{(S_i)_{22}}
\ee
are $\sigma-$independent. They are also equal to each other,  since the $\textrm{det} (S_i)=0$. Therefore we can set
\be
a_i\equiv \frac{(S_i)_{11}}{(S_i)_{12}}=\frac{(S_i)_{21}}{(S_i)_{22}},
\ee
where $a_i$ depends only on $\sigma_0,{\epsilon_0}$ and $\omega$. We can parameterize the  matrices $S_i$
as follows 
\be
\label{param1}
S_i=\begin{pmatrix} a_i  p_i(\sigma) & p_i(\sigma)\\  a_i q_i(\sigma) & q_i(\sigma)\end{pmatrix}.
\ee
Equation \eqref{CaleyHamilton} also completely determines $Y^{-1}(\sigma)$. In fact
\be
\begin{split}
Y^{-1}(\sigma)=&\frac{1}{\textrm{det}(Y)}( \mathrm{tr}(Y){\mathbb{I}_2}-Y(\sigma))=\\
=&\frac{1}{\textrm{det}(Y)}[ ( \mathrm{tr}(S_1) {\mathbb{I}_2}-S_1) 
e^{\omega(\sigma-{\epsilon_0})}+ (\mathrm{tr}(S_2) {\mathbb{I}_2}-S_2) 
e^{-\omega(\sigma-{\epsilon_0})}] \,.
\end{split}
\ee
Next we construct the bilinear $Y^{-1} P_0^{-1} Y$. We find
\begin{eqnarray}
\label{param2}
Y^{-1} P_0^{-1} Y&=\frac{1}{\textrm{det}(Y)}[ ( \mathrm{tr}(S_1) {\mathbb{I}_2}-S_1) 
e^{\omega(\sigma-{\epsilon_0})}+ (\mathrm{tr}(S_2) {\mathbb{I}_2}-S_2) 
e^{-\omega(\sigma-{\epsilon_0})}]P_0^{-1} [e^{\omega(\sigma-{\epsilon_0})}S_1+\nn\\ &+ e^{-\omega(\sigma-{\epsilon_0})}S_2]
\equiv  \mathcal{A}_2 e^{2\omega(\sigma-{\epsilon_0})}+\mathcal{B}_2 e^{-2\omega(\sigma-{\epsilon_0})}+
\mathcal{A}_0+\mathcal{B}_0
\end{eqnarray}
with
\be
\begin{split}
\label{param3}
\mathcal{A}_2=&\frac{1}{\textrm{det}(Y)}( \mathrm{tr}(S_1) {\mathbb{I}_2}-S_1) P_0^{-1} S_1\,,\ \ \ \ 
\mathcal{B}_2=\frac{1}{\textrm{det}(Y)}( \mathrm{tr}(S_2) {\mathbb{I}_2}-S_2) P_0^{-1} S_2\,,\\
\mathcal{A}_0=&\frac{1}{\textrm{det}(Y)}( \mathrm{tr}(S_1) {\mathbb{I}_2}-S_1) P_0^{-1} S_2\,, \ \ \ 
\mathcal{B}_0=\frac{1}{\textrm{det}(Y)}( \mathrm{tr}(S_2) {\mathbb{I}_2}-S_2) P_0^{-1} S_1 \,.
\end{split}
\ee
Because of the relations \eqref{cipro} we find that the matrices $\mathcal{A}_2$  and $\mathcal{B}_2$ are nihilpotent
\be
\mathcal{A}_2^2=\mathcal{B}^2_2=0\,,
\ee
and it holds
\be
\mathcal{A}_2 \mathcal{A}_0=\mathcal{A}_0 \mathcal{B}_2=\mathcal{B}_2 \mathcal{B}_0=
\mathcal{B}_0 \mathcal{A}_2=\mathcal{B}_0 \mathcal{A}_0=\mathcal{A}_0 \mathcal{B}_0=0.
\ee
The structure of the matrices $\mathcal{A}_i$ and $\mathcal{B}_i$ is very simple. They are in fact constant matrices
times a function of $\sigma$. This can be easily shown  by means of the parametrization  \eqref{param1}.  In fact
\begin{align}
\mathcal{A}_2(\sigma)=&\frac{(q_1(\sigma ){}^2-p_1(\sigma ){}^2)}{\textrm{det}(Y)\Lambda(\sigma)}\left(
\begin{array}{cc}
 a_1 & 1 \\
 -a_1^2 & -a_1 \\
\end{array}
\right)\ \   \mathcal{A}_0(\sigma)=\frac{(q_1(\sigma ) q_2(\sigma )-p_1(\sigma ) p_2(\sigma ))}{\textrm{det}(Y)\Lambda(\sigma)} \left(
\begin{array}{cc}
 a_2 & 1 \\
 -a_1 a_2 & -a_1 \\
\end{array}
\right)
\\
\mathcal{B}_2(\sigma)=&\frac{(q_2(\sigma ){}^2-p_2(\sigma ){}^2)}{\textrm{det}(Y)\Lambda(\sigma)}\left(
\begin{array}{cc}
 a_2 & 1 \\
 -a_2^2 & -a_2 \\
\end{array}
\right)\ \   \mathcal{B}_0(\sigma)=\frac{(q_1(\sigma ) q_2(\sigma )-p_1(\sigma ) p_2(\sigma ))}{\textrm{det}(Y)\Lambda(\sigma)} \left(
\begin{array}{cc}
 a_1 & 1 \\
 -a_1 a_2 & -a_2 \\
\end{array}
\right),
\end{align}
where we used that $P_0\equiv\Lambda(\sigma)\sigma_1$.\\
\\
Our next goal is to compute
\be
\textrm{det} \left(\int_{\epsilon_0}^R ds~
 Y^{-1}(s)~ P_0^{-1}(s)~ Y(s)\right)
\ee
in \eqref{op_square_Dirichlet}.
Since $ Y^{-1}(s)~ P_0^{-1}(s)~ Y(s)$ is traceless in our case, we can also write the expression above as 
\be
 =-\frac{1}{2}
 \int_{\epsilon_0}^R\!\!\!\! d\sigma \int_{\epsilon_0}^R\!\!\!\! d\sigma^\prime \mathrm{tr}\left(
 Y^{-1}(\sigma)~ P_0^{-1}(\sigma)~ Y(\sigma)  Y^{-1}(\sigma^\prime)~ P_0^{-1}(\sigma^\prime)~ Y(\sigma^\prime)
 \right)~.
\ee
We can now use the representation  \eqref{param2} and the properties  \eqref{param3} to simplify the expression above.
We obtain
\be
\begin{split}
&\textrm{det} \left(\int_{\epsilon_0}^R ds~
 Y^{-1}(s)~ P_0^{-1}(s)~ Y(s)\right)=\\
 &= (a_1-a_2)^2  \int_{\epsilon_0}^R\!\!\!\! d\sigma\frac{\left(p_1(\sigma ){}^2-q_1(\sigma ){}^2\right)}{\textrm{det}(Y(\sigma))\Lambda(\sigma)}e^{2\omega\sigma} \int_{\epsilon_0}^R\!\!\!\! d\sigma^\prime~\frac{ \left(p_2\left(\sigma
   '\right){}^2-q_2\left(\sigma '\right){}^2\right)}{\textrm{det}(Y(\sigma^\prime))\Lambda(\sigma^\prime)}e^{-2\omega \sigma^\prime}+\\
  &\phantom{=}- (a_1-a_2)^2 \left[\int_{\epsilon_0}^R
 \frac{ \left(p_1(\sigma ) p_2(\sigma )-q_1(\sigma ) q_2(\sigma )\right)}{\textrm{det}(Y(\sigma))\Lambda(\sigma)}\right]^2~.
   \end{split}
\ee
This formula can be very efficiently used to simplify the fermionic determinant in its large-$R$ expansion.
The second line is always negligible, the first one consists of two separate integrals: for positive $\omega$, 
the dominant part in the large-$R$ limit will be the contribution of the first (indefinite) integral evaluated at the upper endpoint times the contribution of the second (indefinite) integral evaluated at the lower endpoint, whereas for negative $\omega$ the roles of first and second integrals are swapped.

\section{Boundary conditions for small Fourier modes}
\label{bclower}

In this Appendix we comment on a different choice for the boundary conditions on the bosonic and fermionic modes with small Fourier mode -- choice followed  in \cite{Kruczenski:2008zk} for the circular Wilson loop case -- and show that it leaves unaffected the main results of this paper, the effective actions for both the circle \eqref{Gammacircular} and the normalized latitude \eqref{discrepancy}.

In \cite{Kruczenski:2008zk} the one-dimensional spectral problems in the radial coordinate $\sigma$ are subject to Dirichlet boundary conditions at both the boundaries  $\sigma=\epsilon_0$ and (fictitious) $\sigma=R$, except for the modes labeled by $m=0$~\footnote{In the labelling of (5.35) after the supersymmetry-preserving regularization.}, for which Neumann boundary conditions are imposed at $\sigma=R$~\footnote{See formulas (5.46)-(5.52) therein.}. 
It is easy to modify our analysis of the bosonic sector in Section \ref{subsec:bosonicdets} -- where we kept Dirichlet boundary conditions for all modes -- and evaluate the effect of this other choice. The relevant Fourier frequency corresponds to $\ell=0$ which, from the discussion at the beginning of Section \ref{sec:partitionfunctions}, corresponds to the mode $\omega=0$ for $\mathcal{O}_1$ and $\mathcal{O}_2(\theta_0)$, $\omega=1$ for $\mathcal{O}_{3+}(\theta_0)$ and $\omega=-1$ for $\mathcal{O}_{3-}(\theta_0)$. We use the subscript $N$ to denote the new determinants with Neumann boundary conditions in $\sigma=R$
\begin{gather}
f_1(\epsilon_0)=\partial_{\sigma} f_1(R)=0\,,
\end{gather}
instead of the Dirichlet ones $f_1(\epsilon_0)=f_1(R)=0$ used in the main text. 
We read off the result of the Gel'fand-Yaglom method for the $\omega=0$ frequency of \eqref{detO1} from (5.48) of \cite{Kruczenski:2008zk}, since the operator $\mathcal{O}_1$ is the same for the circle and the latitude:
\begin{gather}
\left[\textrm{Det}_{\omega=0} \mathcal{O}_1\right]_{N} = \coth\epsilon_0 .
\end{gather}
For the other operators, the new boundary conditions change \eqref{O2first} as 
\begin{gather}
M=
\left(\begin{array}{cc}
1 & 0\\
0 & 0
\end{array}\right)
\qquad
N=
\left(\begin{array}{cc}
0 & 0\\
0 & 1
\end{array}\right)
\qquad
\frac{\textrm{Det}_{\omega}\left[\frac{d^2}{d\sigma^2}+P_2(\sigma)\right]}
{\textrm{Det}_{\omega}\left[\frac{d^2}{d\sigma^2}+\hat{P}_2(\sigma)\right]}
=
\frac{\partial_\sigma f_{(II)1}\left(R\right)}{\partial_\sigma\hat{f}_{(II)1}\left(R\right)}
\end{gather}
and accordingly modify \eqref{detO2},\eqref{detO3plus} and \eqref{detO3minus}  as
\begin{eqnarray}
\left[\textrm{Det}_{\omega=0} \mathcal{O}_2(\theta_0)\right]_{N} &=& \tanh\left(\sigma_0+\epsilon \right)\,, \nonumber
\\
\left[\textrm{Det}_{\omega=1} \mathcal{O}_{3+}(\theta_0)\right]_{N}   &=& \left[\textrm{Det}_{\omega=-1} \mathcal{O}_{3-}(\theta_0)\right]_{N}  = \frac{e^{\sigma_0+2 \epsilon_0 }}{\sqrt{1+e^{2 \sigma_0+4 \epsilon_0 }}} ~.
\end{eqnarray}
The limit $\sigma_0\to\infty$
\begin{gather}
\left[\textrm{Det}_{\omega=0} \mathcal{O}_2(\theta_0=0)\right]_{N} =
\left[\textrm{Det}_{\omega=1} \mathcal{O}_{3+}(\theta_0=0)\right]_{N}  =
\left[\textrm{Det}_{\omega=-1} \mathcal{O}_{3-}(\theta_0=0)\right]_{N}  = 1 
\end{gather}
modifies the analogous results \eqref{detO2cir}-\eqref{detO3minuscir} for the circular Wilson loop. A comparison with the formulas in the main text reveals that, at the level of the Gel'fand-Yaglom determinants, the only change following from this different choice of boundary conditions  is an overall rescaling of the determinants by~$R$.

The same phenomenon occurs in the fermionic sector, where backtracking the special Fourier mode to our $\omega$-labeling is less transparent, but becomes more visible in the circular Wilson loop. The frequency $m=0$ of formula (5.35) \cite{Kruczenski:2008zk} is the determinant of the operator
\begin{gather}\label{sing_op}
-\partial_\sigma^2+\frac{1}{2}+\frac{3}{4 \sinh^2\sigma}-\frac{1}{2} \coth\sigma \, .
\end{gather}
To find it in the present paper, we begin with the Gel'fand-Yaglom differential equation
\begin{gather}
\large[\mathcal{O}_F^{1,1,1}(\theta_0=0)\large]^2
\left(\begin{array}{c}
f_1 (\sigma)\\
f_2 (\sigma)
\end{array}\right)
=
\left(\begin{array}{c}
0\\
0
\end{array}\right) \:
\end{gather}
and from its component equations
\begin{eqnarray}
\left[-\partial_\sigma^2+\left(\omega-\frac{1}{2}\right)^2+\frac{1}{4}+\frac{3}{4 \sinh^2\sigma}-\left(\omega-\frac{1}{2}\right) \coth\sigma \right]f_1 (\sigma)=0 & \\
\left[-\partial_\sigma^2+\left(\omega-\frac{1}{2}\right)^2+\frac{1}{4}+\frac{3}{4 \sinh^2\sigma}+\left(\omega-\frac{1}{2}\right) \coth\sigma \right]f_2 (\sigma)=0 &.
\end{eqnarray}
it is evident that \eqref{sing_op} governs $f_{1}(\sigma)$ for $\omega=1$ while $f_{2}(\sigma)$ for $\omega=0$.\\
\\
Extending this identification to arbitrary $\theta_0$, this argument tells that the only modification in Appendix \ref{sec:square} is the Neumann boundary condition on the first component for $\omega=1$
\begin{gather}
f_1\left({\epsilon_0}\right)=f_2\left({\epsilon_0}\right)=\partial_\sigma f_1\left(R\right)=f_2\left(R\right)=0 \,,
\end{gather}
which translates into replacing \eqref{DDfermMN} with
\begin{gather}
M_{\mathcal{O}^2}=
\left(\begin{array}{cccc}
1 & 0 & 0 & 0\\
0 & 1 & 0 & 0\\
0 & 0 & 0 & 0\\
0 & 0 & 0 & 0\\
\end{array}\right)
\qquad
N_{\mathcal{O}^2}=
\left(\begin{array}{cccc}
0 & 0 & 0 & 0\\
0 & 0 & 0 & 0\\
0 & 0 & 1 & 0\\
0 & 1 & 0 & 0\\
\end{array}\right) \,,
\end{gather}
and on the second component for $\omega=0$
\begin{gather}
f_1\left({\epsilon_0}\right)=f_2\left({\epsilon_0}\right)=f_1\left(R\right)=\partial_\sigma f_2\left(R\right)=0 \,,
\end{gather}
which is implemented by 
\begin{gather}
M_{\mathcal{O}^2}=
\left(\begin{array}{cccc}
1 & 0 & 0 & 0\\
0 & 1 & 0 & 0\\
0 & 0 & 0 & 0\\
0 & 0 & 0 & 0\\
\end{array}\right)
\qquad
N_{\mathcal{O}^2}=
\left(\begin{array}{cccc}
0 & 0 & 0 & 0\\
0 & 0 & 0 & 0\\
1 & 0 & 0 & 0\\
0 & 0 & 0 & 1\\
\end{array}\right) \,.
\end{gather}
This also means that we cannot use the compact form \eqref{op_square_Dirichlet} (still valid for $\omega\neq0,1$) and we have resort to the general expression \eqref{op_square}. After a lengthy computation, the new values of the determinants 
{\small
\begin{align}\nonumber
&
\left[\textrm{Det}_{\omega=1}[(\mathcal{O}_{F}^{{1},1,1})^2]\right]_{N}\!=\!
  \sqrt{R}\,e^{R}\,\frac{e^{-\frac{\sigma_0}{2}} (\tanh\sigma_0+1) \sinh \epsilon_0 \cosh (\sigma_0+\epsilon_0)}{\left(e^{2 \sigma_0}+1\right)^3 \sqrt{2 \cosh (\sigma_0+2 \epsilon_0)}}
  \Big[-2 e^{4\sigma_0} \Big(\log \frac{e^{2 \epsilon_0 }-1}{e^{2 (\sigma_0+\epsilon_0 
)}+1}+
\\ 
&+2 \sigma_0\Big)
+\frac{(e^{2 \sigma _0}+1) \big(e^{6 \sigma _0+4 \epsilon_0 }+(e^{2 \epsilon_0 }+1) e^{4 \sigma _0+2 \epsilon_0 }
  +e^{2 \sigma _0}  (-5 e^{2 \epsilon_0 }+3 e^{4 \epsilon_0 }+3 )+ (e^{2 \epsilon_0 }-1)^2\big)}
  {(e^{2 \epsilon_0 }-1)^2  (e^{2  (\sigma _0+\epsilon_0)}+1)}
\Big]\, 
 \\\nonumber 
 &
\left[\textrm{Det}_{\omega=0}[(\mathcal{O}_{F}^{{1},1,1})^2]\right]_{N}\!=\!
  \sqrt{R}\,e^{R}\,\frac{e^{-\frac{\sigma_0}{2}} (\tanh\sigma_0+1) \sinh \epsilon_0 \cosh (\sigma_0+\epsilon_0)}{\left(e^{2 \sigma_0}+1\right)^2 \sqrt{2 \cosh (\sigma_0+2 \epsilon_0)}}\Big[-2 e^{2 \sigma_0} \Big(\log\frac{e^{2 \epsilon_0 }-1}{e^{2 (\sigma_0+\epsilon_0 )}+1}+\\
&+2 \sigma_0\Big)
+\frac{\left(e^{2\sigma_0}+1\right) \left(-e^{2\sigma_0}+3 e^{2 (\sigma_0+\epsilon_0 )}
+e^{4 (\sigma_0+\epsilon_0 )}-e^{2 \epsilon_0 }+e^{4 \epsilon_0 }+1\right)}{\left(e^{2 \epsilon_0 }-1\right)^2 
\left(e^{2 (\sigma_0+\epsilon_0 )}+1\right)}
\Big]
\end{align}
}
agree with \eqref{specialmode1}-\eqref{specialmode0}, again up to an overall rescaling of their values by a factor of~$\sqrt{R}$.

The analysis in Section \ref{sec:partitionfunctions} goes through in a similar fashion, provided that the lower modes $\Omega^B_{\ell=0} (\theta_0),\,  \Omega^F_{s=\frac{1}{2}} (\theta_0),\, \Omega^F_{s=-\frac{1}{2}} (\theta_0)$ take into account these new determinants. The regularized effective action \eqref{susyreg} does not change when the limit $\mu\to0$ is taken.

\bibliographystyle{nb}
\bibliography{Ref_latitude}

\end{document}